\documentclass[reprint,superscriptaddress,amsmath,amssymb,aps,longbibliography,prb]{revtex4-2}
\usepackage{graphicx}
\usepackage{dcolumn}
\usepackage{bm}
\usepackage{hyperref}
\hypersetup{colorlinks=true,linkcolor=blue,urlcolor=blue,citecolor=blue}
\usepackage{float}
\usepackage{subfigure}
\usepackage{array}
\usepackage{booktabs}
\usepackage{multirow}
\usepackage{CJK}
\usepackage{color}
\usepackage[normalem]{ulem}
\usepackage{soul}  
\usepackage{xcolor}  
\sethlcolor{yellow}
\begin{document}

\title{Chebyshev pseudosite matrix product state approach for cluster perturbation theory}
\author{Pei-Yuan Zhao}
\affiliation{State Key Laboratory of Low Dimensional Quantum Physics and Department of Physics, Tsinghua University, Beijing 100084, China}
\author{Ke Ding}
\affiliation{State Key Laboratory of Low Dimensional Quantum Physics and Department of Physics, Tsinghua University, Beijing 100084, China}
\author{Shuo Yang}
\email{shuoyang@tsinghua.edu.cn}
\affiliation{State Key Laboratory of Low Dimensional Quantum Physics and Department of Physics, Tsinghua University, Beijing 100084, China}
\affiliation{Frontier Science Center for Quantum Information, Beijing 100084, China}
\affiliation{Hefei National Laboratory, Hefei 230088, China}

\begin{abstract}
We introduce the Chebyshev pseudosite matrix product state approach (ChePSMPS) as a solver for cluster perturbation theory (CPT), crucial for simulating spectral functions in two-dimensional electron-phonon ($e$-ph) coupling systems.
ChePSMPS distinguishes itself from conventional exact diagonalization solvers by supporting larger clusters, thereby significantly mitigating finite-size effects.
Free from the fermion sign problem, ChePSMPS enhances its ability to explore $e$-ph effects and generate high-resolution spectral functions in doped Mott insulators.
We use this method to simulate the spectra for both one- and two-dimensional Hubbard-Holstein models, highlighting its superiority over other methods.
Our findings validate ChePSMPS as a powerful and reliable Green's function solver.
In conjunction with embedding methods, ChePSMPS emerges as an essential tool for simulating strongly correlated $e$-ph coupling systems.
\end{abstract}
\maketitle

\section{Introduction}

Spectral functions are pivotal in quantum many-body physics, offering direct insight into the properties of elementary excitations in the system. 
Single-particle spectral functions can be experimentally measured in both momentum space and real space through angle-resolved photoemission spectroscopy (ARPES) and scanning tunneling spectroscopy (STM), respectively. 
These experimental techniques~\cite{RevModPhys.75.473,RevModPhys.79.353} allow direct characterization of electronic properties in actual materials.
The development of effective algorithms for computing spectral functions is crucial, as it bridges theoretical predictions with experimental observations, enhancing our understanding of various intriguing phenomena.

The electron-phonon ($e$-ph) coupling influences the electronic dispersion~\cite{He2018,lanzara2001evidence,Shen2002,PhysRevLett.93.117003,PhysRevLett.95.117001,PhysRevLett.95.227002,PhysRevB.82.064513}, superconducting pairing~\cite{He2018,Shen2002}, thermal Hall effect~\cite{PhysRevLett.124.167601}, pseudogap~\cite{liu2016giant}, and other properties of the cuprates.
Therefore, the study of strongly correlated electronic systems with $e$-ph couplings is of paramount importance. 
However, these systems exhibit high degrees of freedom and pose substantial challenges in computing spectral functions.
The numerical methods fall into two main categories: finite-size and embedding methods. 
Finite-size methods, such as exact diagonalization (ED)~\cite{PhysRevB.69.165115}, determinant quantum Monte Carlo (DQMC)~\cite{PhysRevB.91.165127}, and density matrix renormalization group (DMRG)~\cite{PhysRevB.74.241103}, provide accurate spectral functions but are limited by system size in terms of spectral resolution.
Embedding methods~\cite{sun2016quantum,RevModPhys.68.13,RevModPhys.77.1027}, on the other hand, simulate infinite systems by segmenting them into manageable clusters. 
These methods solve intracluster properties accurately using numerical methods, while treating intercluster correlations approximately as perturbations.
The embedding methods can produce spectral functions with continuous momentum resolution.
In addition, the impact of finite-size effects can be reduced by increasing the size of clusters. 
This paper delves into cluster perturbation theory (CPT)~\cite{PhysRevLett.80.5389,PhysRevLett.84.522,PhysRevB.66.075129}, an embedding method known for its simplicity and effectiveness.

The original CPT algorithm typically employs ED as the solver for the Green's function within the cluster.
However, ED is restricted to small clusters due to computational limitations known as the exponential wall.
Other methods such as ED with optimized boson basis (ED-OBB)~\cite{PhysRevB.71.115201,PhysRevLett.96.156402}, time-dependent DMRG (tDMRG)~\cite{PhysRevB.93.081107}, dynamical DMRG (DDMRG)~\cite{PhysRevB.92.085128}, Lanczos method with matrix product state (MPS)~\cite{paeckel2023matrix}, variational Monte Carlo (VMC)~\cite{PhysRevB.106.245132}, and DQMC~\cite{PhysRevResearch.4.L042015} have also been applied as CPT solvers, each with its own limitations.
Compared to the original ED method, ED-OBB truncates the boson basis, but remains limited by the exponential wall in terms of system size.
tDMRG faces challenges in handling systems with large phonon degrees of freedom, and the computation of wave functions over long times demands substantial computational effort.
DDMRG can simulate fairly accurate spectral functions, but it has a high computational complexity.
The Lanczos MPS method requires an additional reorthogonalization procedure ~\cite{PhysRevB.85.205119,baker2021direct}, which makes it more complex and computationally expensive.
VMC allows for the simulation of large-scale two-dimensional systems with a relatively low computational cost, but its accuracy greatly depends on the choice of effective subspaces.
Furthermore, DQMC encounters difficulties in obtaining spectral functions at low temperatures for models affected by the fermion sign problem~\cite{PhysRevB.87.235133,PhysRevB.96.205141}.

To overcome the constraints of previous methods, we present the recently developed Chebyshev pseudosite matrix product state (ChePSMPS) method~\cite{PhysRevResearch.5.023026}, which merges the merits of the Chebyshev MPS approach~\cite{PhysRevB.83.195115,PhysRevB.91.115144,PhysRevB.92.115130,PhysRevB.97.075111} with the pseudosite DMRG strategy~\cite{PhysRevB.57.6376}, serving as an innovative solver for CPT.
This solver is designed to significantly improve the accuracy and performance of CPT calculations.
We introduce the formulation and implementation of CPT combined with ChePSMPS, subsequently employing this method to simulate the spectral functions of both one-dimensional (1D) and two-dimensional (2D) Hubbard-Holstein models (HHM), thereby demonstrating its advantages over other methods.

\section{CPT method}
\begin{figure}[tbp]
\centering
\includegraphics[width=1.0\columnwidth]{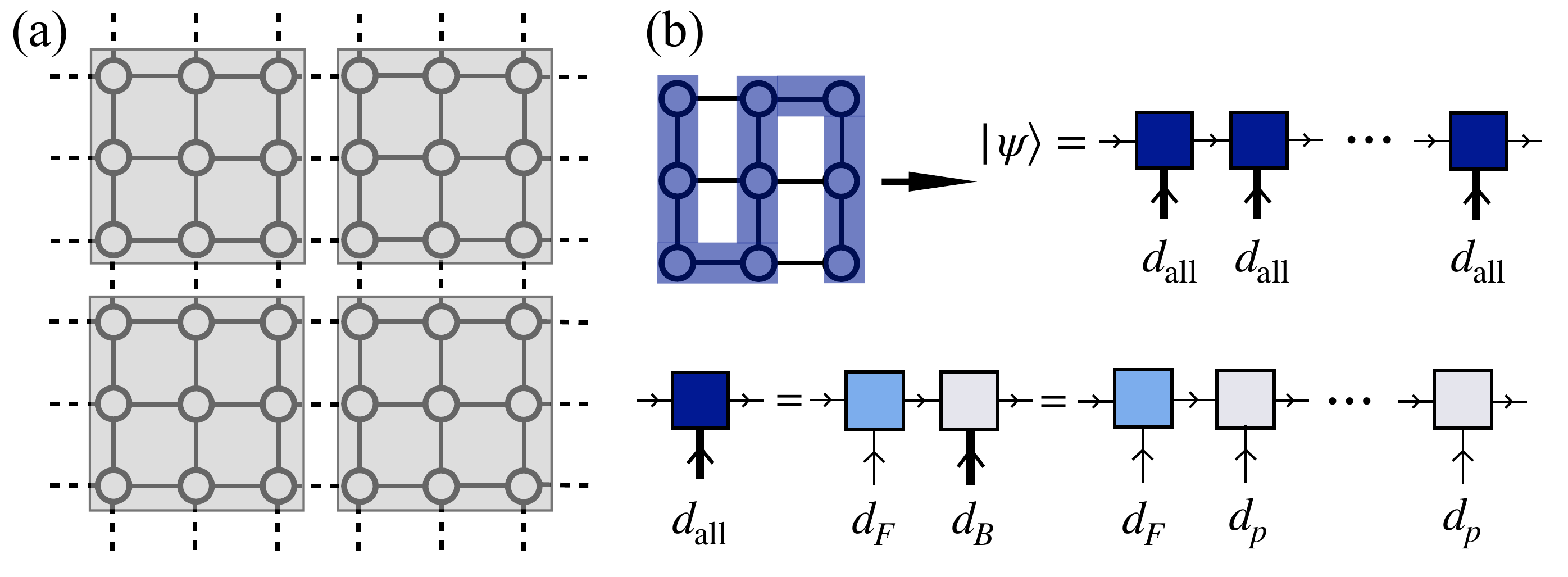}
\caption{
(a) Schematic illustration of the CPT method, where each shaded square represents a cluster and all clusters constitute the entire system.
(b) Pseudosite MPS representation for the wave function within a cluster.
}
\label{Fig.main1}
\end{figure}

In this section, we briefly review the theoretical framework of CPT. 
More detailed derivations and discussions are available in Refs.~\cite{RevModPhys.77.1027,PhysRevLett.84.522,PhysRevB.66.075129}.
We consider a Hamiltonian defined on an infinite lattice with local interactions, where the connections between different clusters are limited to hopping terms.
The system is segmented into a series of clusters, each containing $N$ sites, as shown in Fig.~\ref{Fig.main1}.
The Hamiltonian is decomposed as
\begin{align}
\hat{H}=\sum_{m}\hat{H}_{m}^0+\sum_{m,n\atop i,i'}V_{i,i'}^{m,n}\hat{c}_{m,i}^{\dagger}\hat{c}_{n,i'},
\end{align}
where $\hat{H}_{m}^0$ includes all hopping and interaction terms within the $m$-th cluster.
The intercluster hoppings are written as the hopping matrix $V_{i,i'}^{m,n}$, with $\hat{c}_{m,i}^{\dagger}(\hat{c}_{m,i})$ denoting the fermion creation (annihilation) operators and $i$ representing the site coordinates within a cluster.
The spin index $\sigma$ is omitted for simplicity.

Applying CPT, we treat the intracluster and intercluster terms separately, leading to a mixed representation of the Green's function for the entire system~\cite{PhysRevLett.84.522}
\begin{align}
\mathcal{G}(Q,z)=\bm{G}(z)[\bm{I}-\bm{V}(Q)\bm{G}(z)]^{-1},
\label{Eq_GQz}
\end{align}
where $z=\omega+i\eta$ with $\eta$ denoting the spectral broadening.
$\bm{V}(Q)$ is the Fourier transformation of $V_{i,i'}^{m,n}$ with respect to the cluster coordinates $m$ and $n$.
$Q=(N_xk_x,N_yk_y)$ denotes the superlattice vector, and $\bm{G}(z)$ corresponds to Green's function of the intracluster Hamiltonian $\hat{H}_m$.
The bold symbols in Eq.~(\ref{Eq_GQz}) indicate matrix quantities of dimension $N\times N$, where $N=N_xN_y$ is the size of the cluster.
The Green's function for the entire system is then expressed as
\begin{align}
G(\bm{k},z)=\frac{1}{N}\sum_{i,i'}e^{-i\bm{k}\cdot(\bm{r}_i-\bm{r}_{i'})}\mathcal{G}_{i,i'}(Q,z).
\label{Eq_Gkz}
\end{align}
The imaginary part of this function yields the spectral function
\begin{align}
A(\bm{k},\omega)=-\mathrm{Im} G(\bm{k},z)/\pi.
\end{align}
In the following section, we will illustrate that ChePSMPS~\cite{PhysRevResearch.5.023026} is a suitable solver for CPT.
It can accurately simulate the cluster Green's function at a moderate computational cost and without the fermion sign problem.

\section{CPT + ChePSMPS  implementation}
We now introduce the method that utilizes ChePSMPS as a CPT solver to simulate the spectral function of the strongly correlated system with $e$-ph couplings in the thermodynamic limit, which has three steps:

\begin{figure*}[tbp]
\centering
\includegraphics[width=1\textwidth]{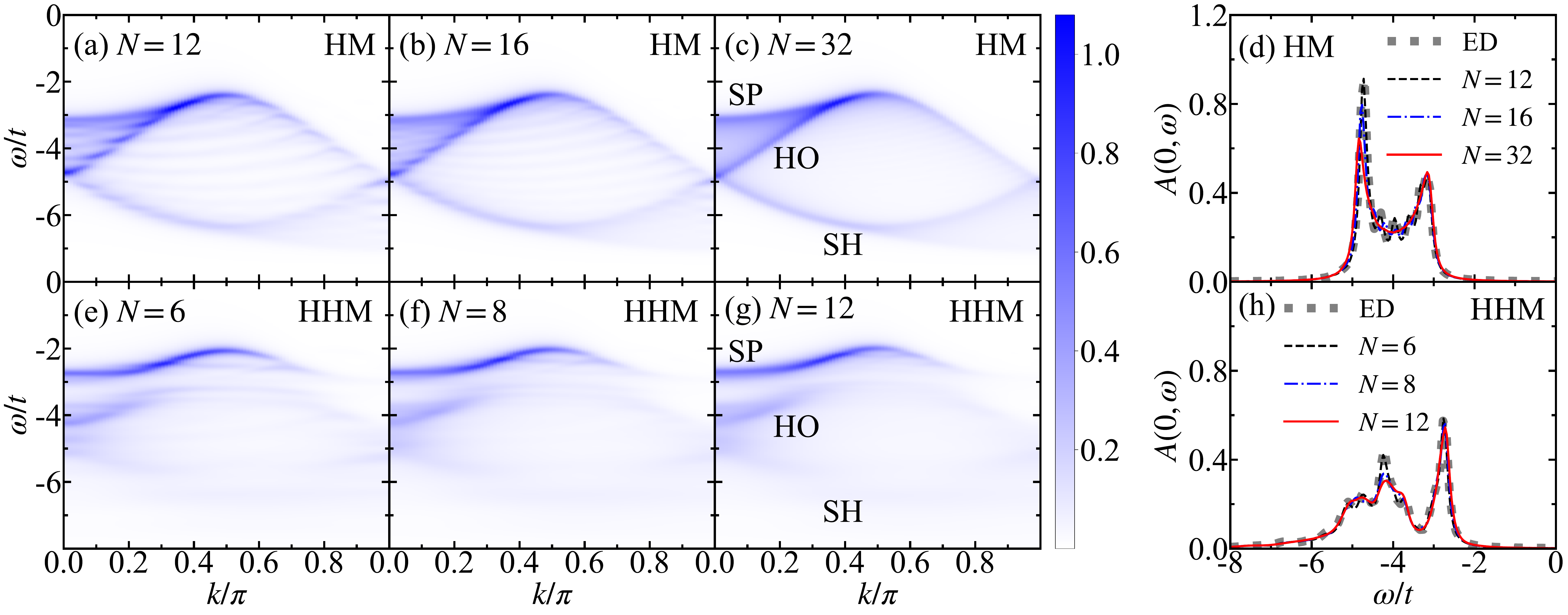}
\caption{
Spectral functions of the 1D half-filled Hubbard(-Holstein) model with $\eta=0.1$, $U/t=8$, $D_{\mathrm{C}}=100$, $N_{\mathrm{C}}=1800$, $\omega_{1\mathrm{max}}=80$, and $\omega_{2\mathrm{max}}=0$.
SP, HO, and SH represent the spinon, holon, and shadow bands, respectively.
(a)-(d) HM with different cluster sizes, with the dotted gray line in (d) representing results from the ED-Lanczos method for $N=12$.
(e)-(h) HHM with different cluster sizes, setting $\omega_0/t=1$, $\gamma/t=1$, and $N_b=7$. 
The dotted gray line in (h) displays results from the ED-Lanczos method with $N=6$ and $N_b=6$.
}
\label{Fig.main2}
\end{figure*}

(1) Map the system within the cluster to a 1D chain. 
Employ the pseudosite matrix product state and matrix product operator (MPO)~\cite{schollwock2011density,PhysRevB.57.6376,PhysRevResearch.5.023026} to describe its wave function and Hamiltonian, respectively. 
As illustrated in Fig.~\ref{Fig.main1}(b), the original MPS at a lattice site is decomposed into fermionic MPS and bosonic MPS, where $d_F$ and $d_B$ represent the fermionic and bosonic degrees of freedom, respectively. 
Here, we choose $d_B=N_b+1=2^{N_p}$, with $N_b$ indicating the maximum number of bosons per lattice site and $N_p$ being an integer.
Subsequently, the bosonic MPS is further decomposed into $N_p=\log_2 d_B$ pseudosite MPSs, each with $d_p=2$ degrees of freedom.  
The total physical degrees of freedom at each lattice site is $d_{\mathrm{all}}=d_F \times d_B$.
Following this segmentation, we employ DMRG~\cite{PhysRevLett.69.2863,PhysRevB.48.10345,PhysRevB.57.6376,PhysRevB.72.180403} to determine the ground state wave function $|\psi_0\rangle$ and its energy $E_0$.

(2) Expand the cluster Green's function using Chebyshev polynomials $T_n(\omega)$ \cite{RevModPhys.78.275}. 
The cluster Green's function is defined as $G_{i,i'}(z)\equiv G_{i,i'}^+(z)-G_{i',i}^{-}(z)$, where
\begin{align}
G_{i,i'}^{\pm}(z)=\langle\psi_0|\hat{c}^{\mp}_{i}\frac{1}{\pm z+(E_0-\hat{H})}\hat{c}_{i'}^{\pm}|\psi_0\rangle,
\label{Eq_Gpmz}
\end{align}
$\hat{c}_i^{+}=\hat{c}_i^{\dagger}$ and $\hat{c}_i^{-}=\hat{c}_i$.
Since the domain of $T_n(\omega)$ is $[-1,1]$, we need to map $\omega$ and $\hat{H}$ to this range
\begin{align}
\omega&\mapsto\omega',\ \omega'=\frac{\omega}{a}+b,\ \omega'\in [-W',W'],\ z'=\omega'+i\eta/a,\nonumber\\
\hat{H}&\mapsto\hat{H'},\ \hat{H'}=\frac{\hat{H}-E_0}{a}+b.
\label{Eq_linear_map}
\end{align}
In Eq.~(\ref{Eq_linear_map}), $W'$ is a number slightly smaller than $1$, the selection of constants $a$ and $b$ refers to Appendix~\ref{AP1}.
Then, we expand $G_{i,i'}(z')$ to obtain
\begin{align}
G_{i,i'}(z')&=\frac{1}{a}\sum_{n=0}^{N_{\mathrm{C}}-1}g_n\big{(}\alpha^+_n(z')\mu_n^+-\alpha^-_n(z')\mu_n^-\big{)},
\label{Eq_Grrpzp}
\end{align}
where $\mu_n^{+}=\langle\psi_0|\hat{c}_{i}T_n(\hat{H}')\hat{c}_{i'}^{\dagger}|\psi_0\rangle$ and $\mu_n^{-}=\langle\psi_0|\hat{c}^{\dagger}_{i'}T_n(\hat{H}')\hat{c}_{i}|\psi_0\rangle$ are the model-dependent expansion coefficients, $N_{\mathrm{C}}$ is the expansion order, $g_n$ is the damping factor for suppressing Gibbs oscillations (here we adopt Jackson damping \cite{RevModPhys.78.275}), and $\alpha^{\pm}_n(z')$ are known functions independent of the model (see Appendix \ref{AP1}).
Therefore, obtaining $\mu_n^{\pm}$ is sufficient to compute $G_{i,i'}(z)$.

(3) Employ the ChePSMPS method to determine $\mu_n^{\pm}$, and then derive the spectral function.
Taking the calculation of $\mu_n^+$ as an example, we use the variational pseudosite MPS method \cite{schollwock2011density,PhysRevB.57.6376,PhysRevResearch.5.023026} to compute each order of Chebyshev vectors,
\begin{align}
|t_0\rangle &=\hat{c}_{i'}^{\dagger}|\psi_0\rangle,\ |t_1\rangle = \hat{H}'|t_0\rangle,\nonumber\\
|t_{n+1}\rangle &=2\hat{H}'|t_n\rangle-|t_{n-1}\rangle,
\label{CheIteration}
\end{align}
thus obtaining $\mu_n^+=\langle\psi_0|\hat{c}_i|t_n\rangle$, and similarly for $\mu_n^-$.
Then, substitute $\mu_n^{\pm}$ to Eq.~(\ref{Eq_Grrpzp}) to compute the cluster Green's function Eq.~(\ref{Eq_Gpmz}).
Next, replace it in Eq.~(\ref{Eq_GQz}) and Eq.~(\ref{Eq_Gkz}) to derive the Green's function $G(\bm{k},z)$ of the entire system, from which the imaginary part yields the single-particle spectral function $A(\bm{k},\omega)$.

\section{Application to Hubbard-Holstein model}
The Hamiltonian of the Hubbard-Holstein model is
\begin{align}
\hat{H}=&-t\sum_{\langle i,j\rangle,\sigma}(\hat{c}_{i,\sigma}^{\dagger}\hat{c}_{j,\sigma}+h.c.)+U\sum_i\hat{n}_{i,\uparrow}\hat{n}_{i,\downarrow} \nonumber\\
& +\omega_0\sum_{i}\hat{a}_i^{\dagger}\hat{a}_i +\gamma\sum_i\hat{n}_i(\hat{a}_i^{\dagger}+\hat{a}_i),
\label{Ham_HHM}
\end{align}
where $\hat{a}_i^{\dagger}(\hat{a}_i)$ denotes the phonon creation (annihilation) operator, $\hat{n}_i=\sum_{\sigma}\hat{c}_{i,\sigma}^{\dagger}\hat{c}_{i,\sigma}$ represents the electron number operator, $U$ stands for the onsite Coulomb repulsion, $\omega_0$ signifies the phonon frequency, and $\gamma$ refers to the $e$-ph coupling parameter.

\begin{figure}[tbp]
\centering
\includegraphics[width=1.0\columnwidth]{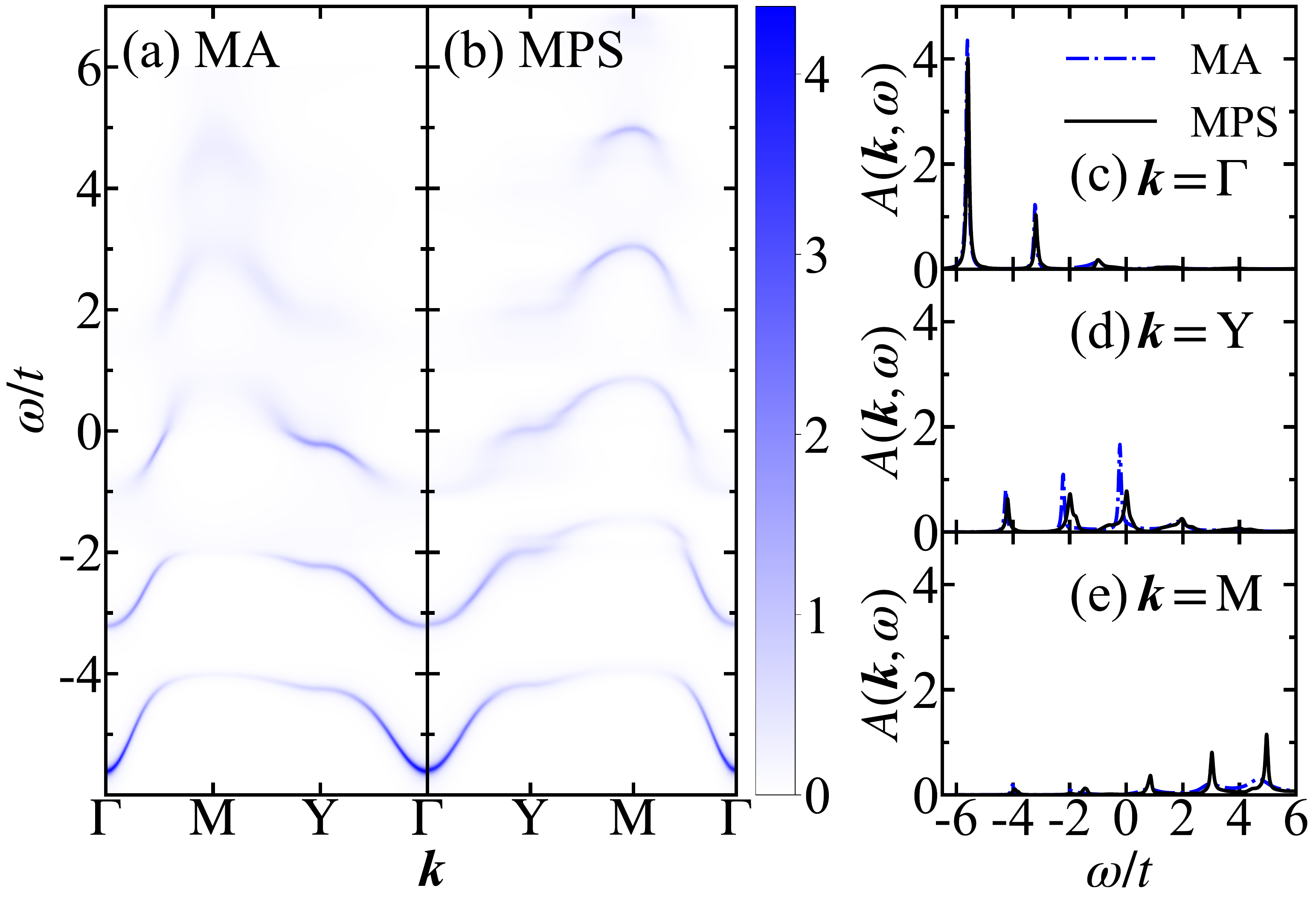}
\caption{Spectra of the Holstein polaron on a square lattice, where $\omega_0/t=2$, $\gamma/t=\sqrt{7.56}$, and $\eta=0.05$. 
High symmetry points $\Gamma$, X, Y, and M correspond to $\bm{k}=(0,0)$, $(\pi,0)$, $(0,\pi)$, and $(\pi,\pi)$, respectively.
(a) The spectrum given by momentum average approximation (MA).
(b) The spectral function calculated using CPT+ChePSMPS, with $N=2\times 12$, $N_b=15$, $N_{\mathrm{C}}=1600$, $D_{\mathrm{C}}=100$,\ $\omega_{1\mathrm{max}}=100$, and $\omega_{2\mathrm{max}}=0$.
(c)-(e) Comparisons of the spectral function calculated by MA and CPT+ChePSMPS at different momenta $\bm{k}$.}
\label{Fig.main3}
\end{figure}

\begin{figure*}[tbp]
\centering
\includegraphics[width=1.8\columnwidth]{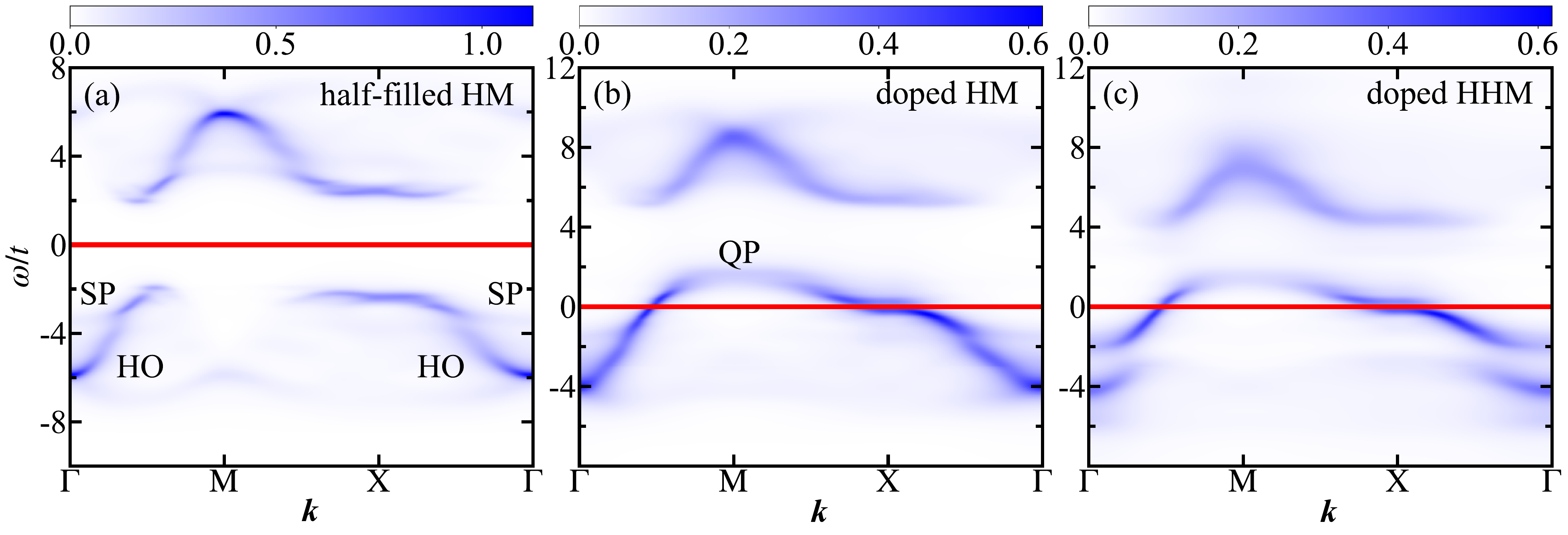}
\caption{Spectra of the (Hubbard-)Holstein model on a square lattice with $N=4\times 4$, $U/t=8$, $\eta=0.1$, and $N_{\mathrm{C}}=400$.
The red line indicates the Fermi level.
SP, HO, and QP denote the spinon-like, holon-like, and quasi-particle bands, respectively. 
(a) Half-filled HM with $N_e=16$ electrons, $D_{\mathrm{C}}=300$, $\omega_{1\mathrm{max}}=80$, and $\omega_{2\mathrm{max}}=0$.
(b) Doped HM with $N_e=14$, $D_{\mathrm{C}}=400$, $\omega_{1\mathrm{max}}=80$, and $\omega_{2\mathrm{max}}=3$.
(c) Doped HHM with $N_e=14$, $N_b=3$, $\omega_0/t=2$, $\gamma/t=\sqrt{3.2}$, $D_{\mathrm{C}}=600$, $\omega_{1\mathrm{max}}=90$, and $\omega_{2\mathrm{max}}=3$.}
\label{Fig.main4}
\end{figure*}

To demonstrate the advantages of our method over traditional CPT+ED, we simulate the spectral functions of the 1D half-filled Hubbard model (HM) and the Hubbard-Holstein model. 
The spinon, holon, and shadow bands for 1D HM are clearly visible in Fig.~\ref{Fig.main2}(a). 
Similarly, spin-charge separation can be clearly observed for HHM in Fig.~\ref{Fig.main2}(e). 
Unlike HM, in HHM the holon band is separated from the spinon band and the shadow band at $k=\pi/2$ and $k=0$, respectively, and the separation approximates the phonon frequency $\omega_0$.
Additionally, phonon interactions significantly reduce the spectral weights of the holon and shadow bands.

We then examine the impact of the cluster size $N$ on the spectrum.
As seen in Figs.~\ref{Fig.main2}(a)-(b) and \ref{Fig.main2}(e)-(f), small clusters result in a spectrum composed of subbands due to finite size effects, which are not inherent to HM and HHM.
Increasing the cluster size almost eliminates these finite size effects, as illustrated in Figs.~\ref{Fig.main2}(c)-(d) for HM and Figs.~\ref{Fig.main2}(g)-(h) for HHM.
Furthermore, our CPT+ChePSMPS results align perfectly with CPT+ED under the same conditions, as shown in Figs.~\ref{Fig.main2}(d) and \ref{Fig.main2}(h).
They are also in full concordance with Fig.~5 in Ref.~\cite{PhysRevLett.96.156402}, confirming the accuracy of our approach.
Traditional ED solvers face limitations due to the exponential wall, restricting them to small clusters. 
This limitation is exacerbated in $e$-ph coupling models by the extensive phonon degrees of freedom. 
As a result, CPT+ED simulations display significant finite size effects.
Our CPT+ChePSMPS approach, however, can handle larger clusters, effectively minimizing finite size effects and yielding higher resolution spectral functions.

To further demonstrate the effectiveness of our method in 2D $e$-ph coupling systems, we simulate spectra of the Holstein polaron,
\begin{align}
A(\bm{k},\omega)=\langle\mathrm{vac}|\hat{c}_{\bm{k}}\delta(\omega-\hat{H})\hat{c}_{\bm{k}}^{\dagger}|\mathrm{vac}\rangle.
\end{align}
Here, $|\mathrm{vac}\rangle$ denotes the vacuum state and $\hat{H}$ is the Hamiltonian of the Holstein model (HHM with $U=0$). 
The subbands depicted in Fig.~\ref{Fig.main3} correspond to an electron band coupled with multiple phonon excitations, and the subband spacing roughly equals the phonon frequency $\omega_0$.
Within the low-energy domain, the CPT+ChePSMPS spectra align well with the momentum average approximation (MA) results~\cite{PhysRevB.74.245104}, as shown in Fig.~\ref{Fig.main3}.
Despite deviations in the high-energy domain compared to MA, our results remain consistent with those obtained by CPT+ED (Fig.~6 in Ref.~\cite{PhysRevB.68.184304}). 
This discrepancy can be attributed to the approximate nature of MA. 
Notably, our ability to simulate larger clusters with CPT+ChePSMPS enhances the resolution of the spectral functions compared to CPT+ED.

Applying CPT+ChePSMPS to strongly correlated 2D systems with $e$-ph interactions, we first examine the spectral function of the HM with $U/t=8$.
Strong Coulomb repulsion manifests itself as the Mott gap, along with the spinon-like and holon-like bands~\cite{kohno2018characteristics}, as evident in Fig.~\ref{Fig.main4}(a).
Upon doping the HM, the Fermi level lowers, perserving the spinon-like and holon-like bands, while introducing a quasi-particle (QP) band~\cite{wang2020emergence} near momentum $\bm{k}=\mathrm{M}$, as shown in Fig.~\ref{Fig.main4}(b).
These Mott characteristics in Figs.~\ref{Fig.main4}(a)-(b) are consistent with prior findings (Fig.~1 in Ref.~\cite{PhysRevLett.108.076401}), confirming the accuracy of our method for 2D doped HM.

Introducing onsite $e$-ph couplings converts HM into HHM, whose zero-temperature spectral function has been relatively unexplored.
As depicted in Fig.~\ref{Fig.main4}(c), the holon-like band divides into two subbands, with a separation close to $\omega_0/t=2$.
These results confirm that CPT+ChePSMPS successfully captures phonon effects in doped Mott insulators. 
The capability of our method to handle larger clusters results in higher-resolution spectral functions compared to that of the ED solver.
Moreover, unlike the DQMC solver, CPT+ChePSMPS is not hindered by the fermion sign problem, making it suitable for investigating $e$-ph coupling effects at zero temperature.

\begin{figure*}[tbp]
\centering
\includegraphics[width=1.8\columnwidth]{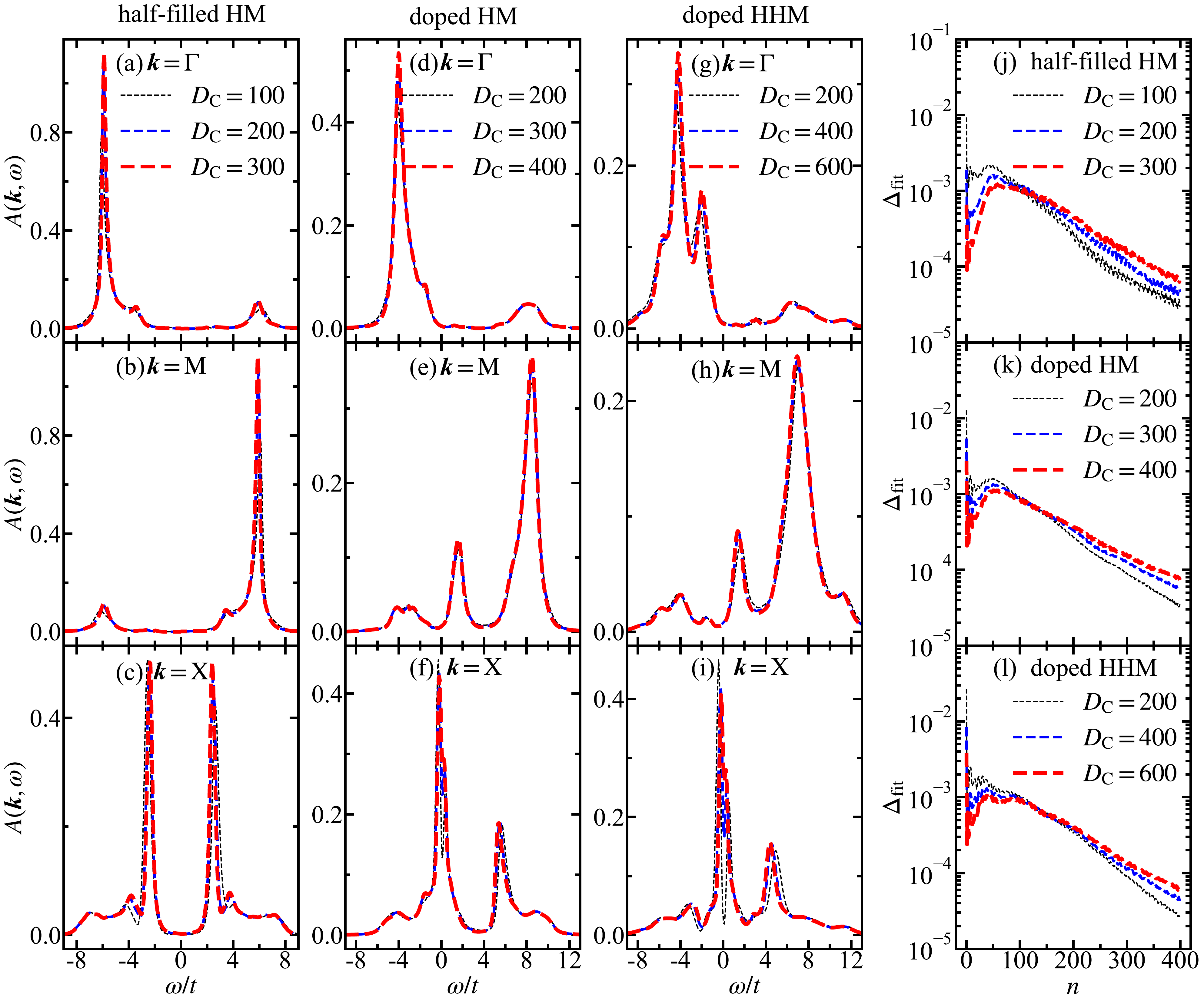}
\caption{
(a)-(c) Convergence of the spectral function $A(\bm{k},\omega)$ at momenta $\bm{k}=\Gamma$, M, and X for the half-filled HM with respect to $D_{\mathrm{C}}$, using the same parameters as in Fig.~\ref{Fig.main4}(a).
(d)-(f) Convergence of $A(\bm{k},\omega)$ at momenta $\bm{k}=\Gamma$, M, and X for the doped HM with respect to $D_{\mathrm{C}}$, using the same parameters as in Fig.~\ref{Fig.main4}(b).
(g)-(i) Convergence of $A(\bm{k},\omega)$ at momenta $\bm{k}=\Gamma$, M, and X for the doped HHM with respect to $D_{\mathrm{C}}$, using the same parameters as in Fig.~\ref{Fig.main4}(c).
(j)-(l) Fitting errors $\Delta_{\mathrm{fit}}$ associated with $G_{1,1}^{-}$ in the half-filled HM, doped HM, and doped HHM, respectively.
}
\label{Fig.main5}
\end{figure*}

\section{Error and convergence Analysis}

This section analyzes the error and convergence properties of the CPT+ChePSMPS method. 
Figures \ref{Fig.main5}(a)-(c) show the convergence of $A(\bm{k},\omega)$ at high symmetry points $\bm{k}=\Gamma$, M, and X for the half-filled HM as a function of $D_{\mathrm{C}}$.  
The spectral function converges at $D_{\mathrm{C}}=300$, with negligible differences from the results at $D_{\mathrm{C}}=100$.
This indicates that the key features of the spectral function are captured at a moderate computational cost. 
In particular, $A(\bm{k}=\Gamma,\omega)$ in Fig.~\ref{Fig.main5}(a) mirrors $A(\bm{k}=\mathrm{M},-\omega)$ in Fig.~\ref{Fig.main5}(b), and $A(\bm{k}=\mathrm{X},\omega)$ equals $A(\bm{k}=\mathrm{X},-\omega)$ in Fig.~\ref{Fig.main5}(c), consisting of the particle-hole symmetry in the half-filled HM and validating the accuracy of our method.
Figures \ref{Fig.main5}(d)-(f) and \ref{Fig.main5}(g)-(i) present the convergence of $A(\bm{k},\omega)$ for the doped HM and the doped HHM as a function of $D_{\mathrm{C}}$. 
HM and HHM exhibit convergence of the spectral function at $D_{\mathrm{C}}=400$ and $D_{\mathrm{C}}=600$, respectively.

In calculating Chebyshev vectors $|t_n\rangle$ in Eq.~(\ref{CheIteration}), we employ a two-site update method ~\cite{schollwock2011density} to variationally minimize the fitting errors
\begin{align}
\Delta_{\mathrm{fit}}=\left\{
\begin{aligned}
&\Vert|t_n\rangle-\hat{c}^{\pm}_i|\psi_0\rangle\Vert^2,\ \ \mathrm{if} \ n =0,\\
&\Vert|t_{n}\rangle-\hat{H}'|t_{n-1}\rangle\Vert^2,\ \ \mathrm{if} \ n =1,\\
&\Vert|t_{n}\rangle-(2\hat{H}'|t_{n-1}\rangle-|t_{n-2}\rangle)\Vert^2,\ \ \mathrm{if} \ n \geq 2.\\
\end{aligned}
\right.
\label{Eq_Deltafit}
\end{align}
For instance, when analyzing the computation of $G_{1,1}^-$, we show the variation of fitting error $\Delta_{\mathrm{fit}}$ with bond dimension $D_{\mathrm{C}}$ and expansion order $n$ for half-filled HM, doped HM, and doped HHM in Figs.~\ref{Fig.main5}(j)-(l).
$\Delta_{\mathrm{fit}}$ is more significant at $n=0$ due to compression of $\hat{c}_1|\psi_0\rangle$ into $|t_0\rangle$, where the large bond dimension $D$ of $|\psi_0\rangle$ contributes to a higher fitting error. 
For $n>1$, $\Delta_{\mathrm{fit}}$ initially increases with $n$ before gradually decreasing. 
In the region where $\Delta_{\mathrm{fit}}$ ascends with $n$, a greater $D_{\mathrm{C}}$ significantly reduces $\Delta_{\mathrm{fit}}$.
In contrast, in the region where $\Delta_{\mathrm{fit}}$ descends with $n$, a larger $D_{\mathrm{C}}$ does not lower $\Delta_{\mathrm{fit}}$ effectively. 
According to the parameters in Fig.~\ref{Fig.main4}, $\Delta_{\mathrm{fit}}$ stays below $10^{-2}$.
Thus, the CPT+ChePSMPS method can capture the essential features of the spectral function with feasible computational effort.

\section{Conclusion}
We present the CPT+ChePSMPS method, which employs ChePSMPS to compute Green's functions within a cluster. 
This approach facilitates the analysis of larger clusters, surpassing the limitations of ED and producing high-solution spectral functions.
When applied to the 2D Holstein polaron Green's function, our findings are consistent with previous studies, underscoring the effectiveness in 2D $e$-ph coupling systems. 
To demonstrate its suitability for strongly correlated cases, we simulate the spectral functions of 2D HM in both half-filled and doped scenarios. 
Our method uncovers Mott features with reasonable computational cost, including Mott gaps, spinon-like bands, and holon-like bands.
Furthermore, we extend to the zero-temperature spectral function of 2D doped HHM, a subject not extensively explored due to the fermion sign problem.
The $e$-ph interactions split the holon-like band into two subbands, separated by roughly the phonon frequency, vividly demonstrating phonon effects.
Our simulations validate ChePSMPS as a powerful solver for Green's function, which, when combined with embedding methods, becomes a valuable tool for investigating 2D strongly correlated electronic systems with $e$-ph couplings. 

\section{Acknowledgement}

This work is supported by the National Natural Science Foundation of China (NSFC) (Grant No. 12174214 and No. 92065205) and the Innovation Program for Quantum Science and Technology (Grant No. 2021ZD0302100).

\appendix

\section{Chebyshev polynomial expansion of the Green's function}
\label{AP1}

In Eq.~(\ref{Eq_Gpmz}), $G^-_{i',i}(z)$ and $G^+_{i,i'}(z)$ include information on the occupied and unoccupied states, respectively.
The energy window for $G_{i,i'}^{+}(z)$ is defined as $\omega \in [-\omega_{2\mathrm{max}}, \omega_{1\mathrm{max}}]$, and for $G_{i',i}^{-}(z)$, it is $\omega \in [-\omega_{1\mathrm{max}}, \omega_{2\mathrm{max}}]$, with the specific choice of $\omega_{1\mathrm{max}}$ and $\omega_{2\mathrm{max}}$ being model-dependent.
Our goal is to expand
\begin{align}
\frac{1}{\pm z-(\hat{H}-E_0)}
\label{Eq_1oz}
\end{align}
using $T_n(\omega)$.
This requires mapping $\omega$ and $\hat{H}$ to the domain of $T_n(\omega) \in [-1,1]$.
For $G^{+}_{i,i'}(z)$, the mapping is
\begin{align}
\omega \mapsto \omega', \ \omega' = \frac{\omega}{a} + b,
\end{align}
and conversely, for $G^{-}_{r',r}(z)$, the mapping is
\begin{align}
-\omega \mapsto -\omega', \ -\omega' = -\frac{\omega}{a} + b.
\end{align}
After mapping, we obtain
\begin{align}
\omega'\in [-W',W'],\ z'=\omega'+i\eta/a.
\end{align}
Similarly, $\hat{H}$ is mapped as
\begin{align}
\hat{H}&\mapsto\hat{H'},\ \hat{H'}=\frac{\hat{H}-E_0}{a}+b.
\end{align}
In these mappings,
\begin{align}
a=\frac{\omega_{1\mathrm{max}}+\omega_{2\mathrm{max}}}{2W'},\ b=\frac{\omega_{2\mathrm{max}}-\omega_{1\mathrm{max}}}{\omega_{2\mathrm{max}}+\omega_{1\mathrm{max}}}W',
\end{align}
where $W'$ is a number slightly less than $1$.
After linear mapping, Eq.~(\ref{Eq_1oz}) can be rewritten as
\begin{align}
\frac{1}{\pm z-(\hat{H}-E_0)}=\frac{1}{a(\pm z'-\hat{H'})},
\end{align}
and then it can be expanded using $T_n(\hat{H}')$ as
\begin{align}
\frac{1}{a(\pm z'-\hat{H'})}=\frac{1}{a}\sum_{n=0}^{\infty}\alpha_n^{\pm}(z')T_n(\hat{H}'),
\label{Eq_KP_Exp}
\end{align}
with the expansion coefficients
\begin{align}
\alpha_n^{\pm}(z)=\frac{2-\delta_{n,0}}{(\pm z)^{n+1}(1+\sqrt{z^2}\frac{\sqrt{z^2-1}}{z^2})^n\sqrt{1-z^{-2}}},
\end{align}
derived from Ref.~\cite{PhysRevB.90.165112}.
Substituting the expansion Eq.~(\ref{Eq_KP_Exp}) into the definition of the cluster Green's function Eq.(\ref{Eq_Gpmz}), we obtain the specific expression for $G_{i,i'}^{\pm}(z)$,
\begin{align}
G^{+}_{i,i'}(z)&=\frac{1}{a}\sum_{n=0}^{\infty}\alpha^+_n(z')\langle\psi_0|\hat{c}_{i}T_n(\hat{H}')\hat{c}_{i'}^{\dagger}|\psi_0\rangle\nonumber\\
&=\frac{1}{a}\sum_{n=0}^{\infty}\alpha^+_n(z')\mu_n^+.
\end{align}
Similarly,
\begin{align}
G^{-}_{i',i}(z)&=\frac{1}{a}\sum_{n=0}^{\infty}\alpha^-_n(z')\mu_n^-.
\end{align}
This leads to the final form of Eq.~(\ref{Eq_Grrpzp}).

\section{The explicit representation of the hopping matrix for the Hubbard-Holstein model}

For nearest-neighbor hoppings between clusters, the one-dimensional hopping matrix has following form
\begin{align}
V_{i, i'}^{m, n}=-t(\delta_{m, n-1} \delta_{i, N} \delta_{i', 1}+\delta_{m, n+1} \delta_{i, 1} \delta_{i', N}),
\label{AP2}
\end{align}
A Fourier transformation on cluster coordinates yields
\begin{align}
V_{i,i'}(Q)=-t(e^{iQ}\delta_{i,N}\delta_{i',1}+e^{-iQ}\delta_{i,1}\delta_{i',N}).
\end{align}
In the case of a two-dimensional square lattice, the hopping matrix is given by
\begin{align}
V_{i,i'}^{m,n}=&-t\big{[}\delta_{(m_x,m_y),(n_x-1,n_y)}\delta_{x,N_x}\delta_{x',1}\delta_{y,y'}\nonumber\\
&+\delta_{(m_x,m_y),(n_x,n_y-1)}\delta_{y,N_y}\delta_{y',1}\delta_{x,x'}\nonumber\\
&+\delta_{(m_x,m_y),(n_x+1,n_y)}\delta_{x,1}\delta_{x',N_x}\delta_{y,y'}\nonumber\\
&+\delta_{(m_x,m_y),(n_x,n_y+1)}\delta_{y,1}\delta_{y',N_y}\delta_{x,x'}\big{]},
\end{align}
where $m=(m_x,m_y)$ and $n=(n_x,n_y)$ are cluster indies.
$i=(x,y)$ and $i'=(x',y')$ denote the coordinates of the lattice points within the cluster.
Applying a Fourier transformation to the cluster indices of $V_{i,i'}^{m,n}$ gives
\begin{align}
V_{i,i'}(Q)=&-t\big{[}e^{iQ_x}\delta_{x,N_x}\delta_{x',1}\delta_{y,y'}\nonumber\\
&+e^{iQ_y}\delta_{y,N_y}\delta_{y',1}\delta_{x,x'}\nonumber\\
&+e^{-iQ_x}\delta_{x,1}\delta_{x',N_x}\delta_{y,y'}\nonumber\\
&+e^{-iQ_y}\delta_{y,1}\delta_{y',N_y}\delta_{x,x'}\big{]}.
\end{align}

\bibliography{apssamp}

\begin{thebibliography}{48}%
\makeatletter
\providecommand \@ifxundefined [1]{%
 \@ifx{#1\undefined}
}%
\providecommand \@ifnum [1]{%
 \ifnum #1\expandafter \@firstoftwo
 \else \expandafter \@secondoftwo
 \fi
}%
\providecommand \@ifx [1]{%
 \ifx #1\expandafter \@firstoftwo
 \else \expandafter \@secondoftwo
 \fi
}%
\providecommand \natexlab [1]{#1}%
\providecommand \enquote  [1]{``#1''}%
\providecommand \bibnamefont  [1]{#1}%
\providecommand \bibfnamefont [1]{#1}%
\providecommand \citenamefont [1]{#1}%
\providecommand \href@noop [0]{\@secondoftwo}%
\providecommand \href [0]{\begingroup \@sanitize@url \@href}%
\providecommand \@href[1]{\@@startlink{#1}\@@href}%
\providecommand \@@href[1]{\endgroup#1\@@endlink}%
\providecommand \@sanitize@url [0]{\catcode `\\12\catcode `\$12\catcode
  `\&12\catcode `\#12\catcode `\^12\catcode `\_12\catcode `\%12\relax}%
\providecommand \@@startlink[1]{}%
\providecommand \@@endlink[0]{}%
\providecommand \url  [0]{\begingroup\@sanitize@url \@url }%
\providecommand \@url [1]{\endgroup\@href {#1}{\urlprefix }}%
\providecommand \urlprefix  [0]{URL }%
\providecommand \Eprint [0]{\href }%
\providecommand \doibase [0]{https://doi.org/}%
\providecommand \selectlanguage [0]{\@gobble}%
\providecommand \bibinfo  [0]{\@secondoftwo}%
\providecommand \bibfield  [0]{\@secondoftwo}%
\providecommand \translation [1]{[#1]}%
\providecommand \BibitemOpen [0]{}%
\providecommand \bibitemStop [0]{}%
\providecommand \bibitemNoStop [0]{.\EOS\space}%
\providecommand \EOS [0]{\spacefactor3000\relax}%
\providecommand \BibitemShut  [1]{\csname bibitem#1\endcsname}%
\let\auto@bib@innerbib\@empty
\bibitem [{\citenamefont {Damascelli}\ \emph {et~al.}(2003)\citenamefont
  {Damascelli}, \citenamefont {Hussain},\ and\ \citenamefont
  {Shen}}]{RevModPhys.75.473}%
  \BibitemOpen
  \bibfield  {author} {\bibinfo {author} {\bibfnamefont {A.}~\bibnamefont
  {Damascelli}}, \bibinfo {author} {\bibfnamefont {Z.}~\bibnamefont
  {Hussain}},\ and\ \bibinfo {author} {\bibfnamefont {Z.-X.}\ \bibnamefont
  {Shen}},\ }\bibfield  {title} {\bibinfo {title} {{Angle-resolved
  photoemission studies of the cuprate superconductors}},\ }\href
  {https://doi.org/10.1103/RevModPhys.75.473} {\bibfield  {journal} {\bibinfo
  {journal} {Rev. Mod. Phys.}\ }\textbf {\bibinfo {volume} {75}},\ \bibinfo
  {pages} {473} (\bibinfo {year} {2003})}\BibitemShut {NoStop}%
\bibitem [{\citenamefont {Fischer}\ \emph {et~al.}(2007)\citenamefont
  {Fischer}, \citenamefont {Kugler}, \citenamefont {Maggio-Aprile},
  \citenamefont {Berthod},\ and\ \citenamefont {Renner}}]{RevModPhys.79.353}%
  \BibitemOpen
  \bibfield  {author} {\bibinfo {author} {\bibfnamefont {O.}~\bibnamefont
  {Fischer}}, \bibinfo {author} {\bibfnamefont {M.}~\bibnamefont {Kugler}},
  \bibinfo {author} {\bibfnamefont {I.}~\bibnamefont {Maggio-Aprile}}, \bibinfo
  {author} {\bibfnamefont {C.}~\bibnamefont {Berthod}},\ and\ \bibinfo {author}
  {\bibfnamefont {C.}~\bibnamefont {Renner}},\ }\bibfield  {title} {\bibinfo
  {title} {{Scanning tunneling spectroscopy of high-temperature
  superconductors}},\ }\href {https://doi.org/10.1103/RevModPhys.79.353}
  {\bibfield  {journal} {\bibinfo  {journal} {Rev. Mod. Phys.}\ }\textbf
  {\bibinfo {volume} {79}},\ \bibinfo {pages} {353} (\bibinfo {year}
  {2007})}\BibitemShut {NoStop}%
\bibitem [{\citenamefont {He}\ \emph {et~al.}(2018)\citenamefont {He},
  \citenamefont {Hashimoto}, \citenamefont {Song}, \citenamefont {Chen},
  \citenamefont {He}, \citenamefont {Vishik}, \citenamefont {Moritz},
  \citenamefont {Lee}, \citenamefont {Nagaosa}, \citenamefont {Zaanen},
  \citenamefont {Devereaux}, \citenamefont {Yoshida}, \citenamefont {Eisaki},
  \citenamefont {Lu},\ and\ \citenamefont {Shen}}]{He2018}%
  \BibitemOpen
  \bibfield  {author} {\bibinfo {author} {\bibfnamefont {Y.}~\bibnamefont
  {He}}, \bibinfo {author} {\bibfnamefont {M.}~\bibnamefont {Hashimoto}},
  \bibinfo {author} {\bibfnamefont {D.}~\bibnamefont {Song}}, \bibinfo {author}
  {\bibfnamefont {S.-D.}\ \bibnamefont {Chen}}, \bibinfo {author}
  {\bibfnamefont {J.}~\bibnamefont {He}}, \bibinfo {author} {\bibfnamefont
  {I.~M.}\ \bibnamefont {Vishik}}, \bibinfo {author} {\bibfnamefont
  {B.}~\bibnamefont {Moritz}}, \bibinfo {author} {\bibfnamefont {D.-H.}\
  \bibnamefont {Lee}}, \bibinfo {author} {\bibfnamefont {N.}~\bibnamefont
  {Nagaosa}}, \bibinfo {author} {\bibfnamefont {J.}~\bibnamefont {Zaanen}},
  \bibinfo {author} {\bibfnamefont {T.~P.}\ \bibnamefont {Devereaux}}, \bibinfo
  {author} {\bibfnamefont {Y.}~\bibnamefont {Yoshida}}, \bibinfo {author}
  {\bibfnamefont {H.}~\bibnamefont {Eisaki}}, \bibinfo {author} {\bibfnamefont
  {D.~H.}\ \bibnamefont {Lu}},\ and\ \bibinfo {author} {\bibfnamefont {Z.-X.}\
  \bibnamefont {Shen}},\ }\bibfield  {title} {\bibinfo {title} {{Rapid change
  of superconductivity and electron-phonon coupling through critical doping in
  Bi-2212}},\ }\href {https://doi.org/10.1126/science.aar3394} {\bibfield
  {journal} {\bibinfo  {journal} {Science}\ }\textbf {\bibinfo {volume}
  {362}},\ \bibinfo {pages} {62} (\bibinfo {year} {2018})}\BibitemShut
  {NoStop}%
\bibitem [{\citenamefont {Lanzara}\ \emph {et~al.}(2001)\citenamefont
  {Lanzara}, \citenamefont {Bogdanov}, \citenamefont {Zhou}, \citenamefont
  {Kellar}, \citenamefont {Feng}, \citenamefont {Lu}, \citenamefont {Yoshida},
  \citenamefont {Eisaki}, \citenamefont {Fujimori}, \citenamefont {Kishio}
  \emph {et~al.}}]{lanzara2001evidence}%
  \BibitemOpen
  \bibfield  {author} {\bibinfo {author} {\bibfnamefont {A.}~\bibnamefont
  {Lanzara}}, \bibinfo {author} {\bibfnamefont {P.}~\bibnamefont {Bogdanov}},
  \bibinfo {author} {\bibfnamefont {X.}~\bibnamefont {Zhou}}, \bibinfo {author}
  {\bibfnamefont {S.}~\bibnamefont {Kellar}}, \bibinfo {author} {\bibfnamefont
  {D.}~\bibnamefont {Feng}}, \bibinfo {author} {\bibfnamefont {E.}~\bibnamefont
  {Lu}}, \bibinfo {author} {\bibfnamefont {T.}~\bibnamefont {Yoshida}},
  \bibinfo {author} {\bibfnamefont {H.}~\bibnamefont {Eisaki}}, \bibinfo
  {author} {\bibfnamefont {A.}~\bibnamefont {Fujimori}}, \bibinfo {author}
  {\bibfnamefont {K.}~\bibnamefont {Kishio}}, \emph {et~al.},\ }\bibfield
  {title} {\bibinfo {title} {{Evidence for ubiquitous strong electron--phonon
  coupling in high-temperature superconductors}},\ }\href
  {https://doi.org/https://doi.org/10.1038/35087518} {\bibfield  {journal}
  {\bibinfo  {journal} {Nature}\ }\textbf {\bibinfo {volume} {412}},\ \bibinfo
  {pages} {510} (\bibinfo {year} {2001})}\BibitemShut {NoStop}%
\bibitem [{\citenamefont {Shen}\ \emph {et~al.}(2002)\citenamefont {Shen},
  \citenamefont {Lanzara}, \citenamefont {Ishihara},\ and\ \citenamefont
  {Nagaosa}}]{Shen2002}%
  \BibitemOpen
  \bibfield  {author} {\bibinfo {author} {\bibfnamefont {Z.-X.}\ \bibnamefont
  {Shen}}, \bibinfo {author} {\bibfnamefont {A.}~\bibnamefont {Lanzara}},
  \bibinfo {author} {\bibfnamefont {S.}~\bibnamefont {Ishihara}},\ and\
  \bibinfo {author} {\bibfnamefont {N.}~\bibnamefont {Nagaosa}},\ }\bibfield
  {title} {\bibinfo {title} {{Role of the electron-phonon interaction in the
  strongly correlated cuprate superconductors}},\ }\href
  {https://doi.org/10.1080/13642810208220725} {\bibfield  {journal} {\bibinfo
  {journal} {Philosophical Magazine B}\ }\textbf {\bibinfo {volume} {82}},\
  \bibinfo {pages} {1349} (\bibinfo {year} {2002})}\BibitemShut {NoStop}%
\bibitem [{\citenamefont {Cuk}\ \emph {et~al.}(2004)\citenamefont {Cuk},
  \citenamefont {Baumberger}, \citenamefont {Lu}, \citenamefont {Ingle},
  \citenamefont {Zhou}, \citenamefont {Eisaki}, \citenamefont {Kaneko},
  \citenamefont {Hussain}, \citenamefont {Devereaux}, \citenamefont {Nagaosa},\
  and\ \citenamefont {Shen}}]{PhysRevLett.93.117003}%
  \BibitemOpen
  \bibfield  {author} {\bibinfo {author} {\bibfnamefont {T.}~\bibnamefont
  {Cuk}}, \bibinfo {author} {\bibfnamefont {F.}~\bibnamefont {Baumberger}},
  \bibinfo {author} {\bibfnamefont {D.~H.}\ \bibnamefont {Lu}}, \bibinfo
  {author} {\bibfnamefont {N.}~\bibnamefont {Ingle}}, \bibinfo {author}
  {\bibfnamefont {X.~J.}\ \bibnamefont {Zhou}}, \bibinfo {author}
  {\bibfnamefont {H.}~\bibnamefont {Eisaki}}, \bibinfo {author} {\bibfnamefont
  {N.}~\bibnamefont {Kaneko}}, \bibinfo {author} {\bibfnamefont
  {Z.}~\bibnamefont {Hussain}}, \bibinfo {author} {\bibfnamefont {T.~P.}\
  \bibnamefont {Devereaux}}, \bibinfo {author} {\bibfnamefont {N.}~\bibnamefont
  {Nagaosa}},\ and\ \bibinfo {author} {\bibfnamefont {Z.-X.}\ \bibnamefont
  {Shen}},\ }\bibfield  {title} {\bibinfo {title} {{Coupling of the ${B}_{1g}$
  Phonon to the Antinodal Electronic States of
  ${\mathrm{B}\mathrm{i}}_{2}{\mathrm{S}\mathrm{r}}_{2}{\mathrm{C}\mathrm{a}}_{0.92}{\mathrm{Y}}_{0.08}{\mathrm{C}\mathrm{u}}_{2}{\mathrm{O}}_{8+\ensuremath{\delta}}$}},\
  }\href {https://doi.org/10.1103/PhysRevLett.93.117003} {\bibfield  {journal}
  {\bibinfo  {journal} {Phys. Rev. Lett.}\ }\textbf {\bibinfo {volume} {93}},\
  \bibinfo {pages} {117003} (\bibinfo {year} {2004})}\BibitemShut {NoStop}%
\bibitem [{\citenamefont {Zhou}\ \emph {et~al.}(2005)\citenamefont {Zhou},
  \citenamefont {Shi}, \citenamefont {Yoshida}, \citenamefont {Cuk},
  \citenamefont {Yang}, \citenamefont {Brouet}, \citenamefont {Nakamura},
  \citenamefont {Mannella}, \citenamefont {Komiya}, \citenamefont {Ando},
  \citenamefont {Zhou}, \citenamefont {Ti}, \citenamefont {Xiong},
  \citenamefont {Zhao}, \citenamefont {Sasagawa}, \citenamefont {Kakeshita},
  \citenamefont {Eisaki}, \citenamefont {Uchida}, \citenamefont {Fujimori},
  \citenamefont {Zhang}, \citenamefont {Plummer}, \citenamefont {Laughlin},
  \citenamefont {Hussain},\ and\ \citenamefont {Shen}}]{PhysRevLett.95.117001}%
  \BibitemOpen
  \bibfield  {author} {\bibinfo {author} {\bibfnamefont {X.~J.}\ \bibnamefont
  {Zhou}}, \bibinfo {author} {\bibfnamefont {J.}~\bibnamefont {Shi}}, \bibinfo
  {author} {\bibfnamefont {T.}~\bibnamefont {Yoshida}}, \bibinfo {author}
  {\bibfnamefont {T.}~\bibnamefont {Cuk}}, \bibinfo {author} {\bibfnamefont
  {W.~L.}\ \bibnamefont {Yang}}, \bibinfo {author} {\bibfnamefont
  {V.}~\bibnamefont {Brouet}}, \bibinfo {author} {\bibfnamefont
  {J.}~\bibnamefont {Nakamura}}, \bibinfo {author} {\bibfnamefont
  {N.}~\bibnamefont {Mannella}}, \bibinfo {author} {\bibfnamefont
  {S.}~\bibnamefont {Komiya}}, \bibinfo {author} {\bibfnamefont
  {Y.}~\bibnamefont {Ando}}, \bibinfo {author} {\bibfnamefont {F.}~\bibnamefont
  {Zhou}}, \bibinfo {author} {\bibfnamefont {W.~X.}\ \bibnamefont {Ti}},
  \bibinfo {author} {\bibfnamefont {J.~W.}\ \bibnamefont {Xiong}}, \bibinfo
  {author} {\bibfnamefont {Z.~X.}\ \bibnamefont {Zhao}}, \bibinfo {author}
  {\bibfnamefont {T.}~\bibnamefont {Sasagawa}}, \bibinfo {author}
  {\bibfnamefont {T.}~\bibnamefont {Kakeshita}}, \bibinfo {author}
  {\bibfnamefont {H.}~\bibnamefont {Eisaki}}, \bibinfo {author} {\bibfnamefont
  {S.}~\bibnamefont {Uchida}}, \bibinfo {author} {\bibfnamefont
  {A.}~\bibnamefont {Fujimori}}, \bibinfo {author} {\bibfnamefont
  {Z.}~\bibnamefont {Zhang}}, \bibinfo {author} {\bibfnamefont {E.~W.}\
  \bibnamefont {Plummer}}, \bibinfo {author} {\bibfnamefont {R.~B.}\
  \bibnamefont {Laughlin}}, \bibinfo {author} {\bibfnamefont {Z.}~\bibnamefont
  {Hussain}},\ and\ \bibinfo {author} {\bibfnamefont {Z.-X.}\ \bibnamefont
  {Shen}},\ }\bibfield  {title} {\bibinfo {title} {{Multiple Bosonic Mode
  Coupling in the Electron Self-Energy of
  $({\mathrm{La}}_{2\ensuremath{-}x}{\mathrm{Sr}}_{x}){\mathrm{CuO}}_{4}$}},\
  }\href {https://doi.org/10.1103/PhysRevLett.95.117001} {\bibfield  {journal}
  {\bibinfo  {journal} {Phys. Rev. Lett.}\ }\textbf {\bibinfo {volume} {95}},\
  \bibinfo {pages} {117001} (\bibinfo {year} {2005})}\BibitemShut {NoStop}%
\bibitem [{\citenamefont {R\"osch}\ \emph {et~al.}(2005)\citenamefont
  {R\"osch}, \citenamefont {Gunnarsson}, \citenamefont {Zhou}, \citenamefont
  {Yoshida}, \citenamefont {Sasagawa}, \citenamefont {Fujimori}, \citenamefont
  {Hussain}, \citenamefont {Shen},\ and\ \citenamefont
  {Uchida}}]{PhysRevLett.95.227002}%
  \BibitemOpen
  \bibfield  {author} {\bibinfo {author} {\bibfnamefont {O.}~\bibnamefont
  {R\"osch}}, \bibinfo {author} {\bibfnamefont {O.}~\bibnamefont {Gunnarsson}},
  \bibinfo {author} {\bibfnamefont {X.~J.}\ \bibnamefont {Zhou}}, \bibinfo
  {author} {\bibfnamefont {T.}~\bibnamefont {Yoshida}}, \bibinfo {author}
  {\bibfnamefont {T.}~\bibnamefont {Sasagawa}}, \bibinfo {author}
  {\bibfnamefont {A.}~\bibnamefont {Fujimori}}, \bibinfo {author}
  {\bibfnamefont {Z.}~\bibnamefont {Hussain}}, \bibinfo {author} {\bibfnamefont
  {Z.-X.}\ \bibnamefont {Shen}},\ and\ \bibinfo {author} {\bibfnamefont
  {S.}~\bibnamefont {Uchida}},\ }\bibfield  {title} {\bibinfo {title}
  {{Polaronic Behavior of Undoped High-${T}_{c}$ Cuprate Superconductors from
  Angle-Resolved Photoemission Spectra}},\ }\href
  {https://doi.org/10.1103/PhysRevLett.95.227002} {\bibfield  {journal}
  {\bibinfo  {journal} {Phys. Rev. Lett.}\ }\textbf {\bibinfo {volume} {95}},\
  \bibinfo {pages} {227002} (\bibinfo {year} {2005})}\BibitemShut {NoStop}%
\bibitem [{\citenamefont {Johnston}\ \emph {et~al.}(2010)\citenamefont
  {Johnston}, \citenamefont {Vernay}, \citenamefont {Moritz}, \citenamefont
  {Shen}, \citenamefont {Nagaosa}, \citenamefont {Zaanen},\ and\ \citenamefont
  {Devereaux}}]{PhysRevB.82.064513}%
  \BibitemOpen
  \bibfield  {author} {\bibinfo {author} {\bibfnamefont {S.}~\bibnamefont
  {Johnston}}, \bibinfo {author} {\bibfnamefont {F.}~\bibnamefont {Vernay}},
  \bibinfo {author} {\bibfnamefont {B.}~\bibnamefont {Moritz}}, \bibinfo
  {author} {\bibfnamefont {Z.-X.}\ \bibnamefont {Shen}}, \bibinfo {author}
  {\bibfnamefont {N.}~\bibnamefont {Nagaosa}}, \bibinfo {author} {\bibfnamefont
  {J.}~\bibnamefont {Zaanen}},\ and\ \bibinfo {author} {\bibfnamefont {T.~P.}\
  \bibnamefont {Devereaux}},\ }\bibfield  {title} {\bibinfo {title}
  {{Systematic study of electron-phonon coupling to oxygen modes across the
  cuprates}},\ }\href {https://doi.org/10.1103/PhysRevB.82.064513} {\bibfield
  {journal} {\bibinfo  {journal} {Phys. Rev. B}\ }\textbf {\bibinfo {volume}
  {82}},\ \bibinfo {pages} {064513} (\bibinfo {year} {2010})}\BibitemShut
  {NoStop}%
\bibitem [{\citenamefont {Chen}\ \emph {et~al.}(2020)\citenamefont {Chen},
  \citenamefont {Kivelson},\ and\ \citenamefont
  {Sun}}]{PhysRevLett.124.167601}%
  \BibitemOpen
  \bibfield  {author} {\bibinfo {author} {\bibfnamefont {J.-Y.}\ \bibnamefont
  {Chen}}, \bibinfo {author} {\bibfnamefont {S.~A.}\ \bibnamefont {Kivelson}},\
  and\ \bibinfo {author} {\bibfnamefont {X.-Q.}\ \bibnamefont {Sun}},\
  }\bibfield  {title} {\bibinfo {title} {{Enhanced Thermal Hall Effect in
  Nearly Ferroelectric Insulators}},\ }\href
  {https://doi.org/10.1103/PhysRevLett.124.167601} {\bibfield  {journal}
  {\bibinfo  {journal} {Phys. Rev. Lett.}\ }\textbf {\bibinfo {volume} {124}},\
  \bibinfo {pages} {167601} (\bibinfo {year} {2020})}\BibitemShut {NoStop}%
\bibitem [{\citenamefont {Liu}\ \emph {et~al.}(2016)\citenamefont {Liu},
  \citenamefont {Konik}, \citenamefont {Rice},\ and\ \citenamefont
  {Zhang}}]{liu2016giant}%
  \BibitemOpen
  \bibfield  {author} {\bibinfo {author} {\bibfnamefont {Y.-H.}\ \bibnamefont
  {Liu}}, \bibinfo {author} {\bibfnamefont {R.~M.}\ \bibnamefont {Konik}},
  \bibinfo {author} {\bibfnamefont {T.}~\bibnamefont {Rice}},\ and\ \bibinfo
  {author} {\bibfnamefont {F.-C.}\ \bibnamefont {Zhang}},\ }\bibfield  {title}
  {\bibinfo {title} {{Giant phonon anomaly associated with superconducting
  fluctuations in the pseudogap phase of cuprates}},\ }\href
  {https://doi.org/https://doi.org/10.1038/ncomms10378} {\bibfield  {journal}
  {\bibinfo  {journal} {Nat. Commun.}\ }\textbf {\bibinfo {volume} {7}},\
  \bibinfo {pages} {10378} (\bibinfo {year} {2016})}\BibitemShut {NoStop}%
\bibitem [{\citenamefont {Fehske}\ \emph {et~al.}(2004)\citenamefont {Fehske},
  \citenamefont {Wellein}, \citenamefont {Hager}, \citenamefont {Wei\ss{}e},\
  and\ \citenamefont {Bishop}}]{PhysRevB.69.165115}%
  \BibitemOpen
  \bibfield  {author} {\bibinfo {author} {\bibfnamefont {H.}~\bibnamefont
  {Fehske}}, \bibinfo {author} {\bibfnamefont {G.}~\bibnamefont {Wellein}},
  \bibinfo {author} {\bibfnamefont {G.}~\bibnamefont {Hager}}, \bibinfo
  {author} {\bibfnamefont {A.}~\bibnamefont {Wei\ss{}e}},\ and\ \bibinfo
  {author} {\bibfnamefont {A.~R.}\ \bibnamefont {Bishop}},\ }\bibfield  {title}
  {\bibinfo {title} {{Quantum lattice dynamical effects on single-particle
  excitations in one-dimensional Mott and Peierls insulators}},\ }\href
  {https://doi.org/10.1103/PhysRevB.69.165115} {\bibfield  {journal} {\bibinfo
  {journal} {Phys. Rev. B}\ }\textbf {\bibinfo {volume} {69}},\ \bibinfo
  {pages} {165115} (\bibinfo {year} {2004})}\BibitemShut {NoStop}%
\bibitem [{\citenamefont {Nowadnick}\ \emph {et~al.}(2015)\citenamefont
  {Nowadnick}, \citenamefont {Johnston}, \citenamefont {Moritz},\ and\
  \citenamefont {Devereaux}}]{PhysRevB.91.165127}%
  \BibitemOpen
  \bibfield  {author} {\bibinfo {author} {\bibfnamefont {E.~A.}\ \bibnamefont
  {Nowadnick}}, \bibinfo {author} {\bibfnamefont {S.}~\bibnamefont {Johnston}},
  \bibinfo {author} {\bibfnamefont {B.}~\bibnamefont {Moritz}},\ and\ \bibinfo
  {author} {\bibfnamefont {T.~P.}\ \bibnamefont {Devereaux}},\ }\bibfield
  {title} {\bibinfo {title} {{Renormalization of spectra by phase competition
  in the half-filled Hubbard-Holstein model}},\ }\href
  {https://doi.org/10.1103/PhysRevB.91.165127} {\bibfield  {journal} {\bibinfo
  {journal} {Phys. Rev. B}\ }\textbf {\bibinfo {volume} {91}},\ \bibinfo
  {pages} {165127} (\bibinfo {year} {2015})}\BibitemShut {NoStop}%
\bibitem [{\citenamefont {Matsueda}\ \emph {et~al.}(2006)\citenamefont
  {Matsueda}, \citenamefont {Tohyama},\ and\ \citenamefont
  {Maekawa}}]{PhysRevB.74.241103}%
  \BibitemOpen
  \bibfield  {author} {\bibinfo {author} {\bibfnamefont {H.}~\bibnamefont
  {Matsueda}}, \bibinfo {author} {\bibfnamefont {T.}~\bibnamefont {Tohyama}},\
  and\ \bibinfo {author} {\bibfnamefont {S.}~\bibnamefont {Maekawa}},\
  }\bibfield  {title} {\bibinfo {title} {{Electron-phonon coupling and
  spin-charge separation in one-dimensional Mott insulators}},\ }\href
  {https://doi.org/10.1103/PhysRevB.74.241103} {\bibfield  {journal} {\bibinfo
  {journal} {Phys. Rev. B}\ }\textbf {\bibinfo {volume} {74}},\ \bibinfo
  {pages} {241103} (\bibinfo {year} {2006})}\BibitemShut {NoStop}%
\bibitem [{\citenamefont {Sun}\ and\ \citenamefont
  {Chan}(2016)}]{sun2016quantum}%
  \BibitemOpen
  \bibfield  {author} {\bibinfo {author} {\bibfnamefont {Q.}~\bibnamefont
  {Sun}}\ and\ \bibinfo {author} {\bibfnamefont {G.~K.-L.}\ \bibnamefont
  {Chan}},\ }\bibfield  {title} {\bibinfo {title} {Quantum embedding
  theories},\ }\href
  {https://doi.org/https://doi.org/10.1021/acs.accounts.6b00356} {\bibfield
  {journal} {\bibinfo  {journal} {Acc. Chem. Res.}\ }\textbf {\bibinfo {volume}
  {49}},\ \bibinfo {pages} {2705} (\bibinfo {year} {2016})}\BibitemShut
  {NoStop}%
\bibitem [{\citenamefont {Georges}\ \emph {et~al.}(1996)\citenamefont
  {Georges}, \citenamefont {Kotliar}, \citenamefont {Krauth},\ and\
  \citenamefont {Rozenberg}}]{RevModPhys.68.13}%
  \BibitemOpen
  \bibfield  {author} {\bibinfo {author} {\bibfnamefont {A.}~\bibnamefont
  {Georges}}, \bibinfo {author} {\bibfnamefont {G.}~\bibnamefont {Kotliar}},
  \bibinfo {author} {\bibfnamefont {W.}~\bibnamefont {Krauth}},\ and\ \bibinfo
  {author} {\bibfnamefont {M.~J.}\ \bibnamefont {Rozenberg}},\ }\bibfield
  {title} {\bibinfo {title} {{Dynamical mean-field theory of strongly
  correlated fermion systems and the limit of infinite dimensions}},\ }\href
  {https://doi.org/10.1103/RevModPhys.68.13} {\bibfield  {journal} {\bibinfo
  {journal} {Rev. Mod. Phys.}\ }\textbf {\bibinfo {volume} {68}},\ \bibinfo
  {pages} {13} (\bibinfo {year} {1996})}\BibitemShut {NoStop}%
\bibitem [{\citenamefont {Maier}\ \emph {et~al.}(2005)\citenamefont {Maier},
  \citenamefont {Jarrell}, \citenamefont {Pruschke},\ and\ \citenamefont
  {Hettler}}]{RevModPhys.77.1027}%
  \BibitemOpen
  \bibfield  {author} {\bibinfo {author} {\bibfnamefont {T.}~\bibnamefont
  {Maier}}, \bibinfo {author} {\bibfnamefont {M.}~\bibnamefont {Jarrell}},
  \bibinfo {author} {\bibfnamefont {T.}~\bibnamefont {Pruschke}},\ and\
  \bibinfo {author} {\bibfnamefont {M.~H.}\ \bibnamefont {Hettler}},\
  }\bibfield  {title} {\bibinfo {title} {{Quantum cluster theories}},\ }\href
  {https://doi.org/10.1103/RevModPhys.77.1027} {\bibfield  {journal} {\bibinfo
  {journal} {Rev. Mod. Phys.}\ }\textbf {\bibinfo {volume} {77}},\ \bibinfo
  {pages} {1027} (\bibinfo {year} {2005})}\BibitemShut {NoStop}%
\bibitem [{\citenamefont {Pairault}\ \emph {et~al.}(1998)\citenamefont
  {Pairault}, \citenamefont {S\'en\'echal},\ and\ \citenamefont
  {Tremblay}}]{PhysRevLett.80.5389}%
  \BibitemOpen
  \bibfield  {author} {\bibinfo {author} {\bibfnamefont {S.}~\bibnamefont
  {Pairault}}, \bibinfo {author} {\bibfnamefont {D.}~\bibnamefont
  {S\'en\'echal}},\ and\ \bibinfo {author} {\bibfnamefont {A.-M.~S.}\
  \bibnamefont {Tremblay}},\ }\bibfield  {title} {\bibinfo {title}
  {{Strong-Coupling Expansion for the Hubbard Model}},\ }\href
  {https://doi.org/10.1103/PhysRevLett.80.5389} {\bibfield  {journal} {\bibinfo
   {journal} {Phys. Rev. Lett.}\ }\textbf {\bibinfo {volume} {80}},\ \bibinfo
  {pages} {5389} (\bibinfo {year} {1998})}\BibitemShut {NoStop}%
\bibitem [{\citenamefont {S\'en\'echal}\ \emph {et~al.}(2000)\citenamefont
  {S\'en\'echal}, \citenamefont {Perez},\ and\ \citenamefont
  {Pioro-Ladri\`ere}}]{PhysRevLett.84.522}%
  \BibitemOpen
  \bibfield  {author} {\bibinfo {author} {\bibfnamefont {D.}~\bibnamefont
  {S\'en\'echal}}, \bibinfo {author} {\bibfnamefont {D.}~\bibnamefont
  {Perez}},\ and\ \bibinfo {author} {\bibfnamefont {M.}~\bibnamefont
  {Pioro-Ladri\`ere}},\ }\bibfield  {title} {\bibinfo {title} {{Spectral Weight
  of the Hubbard Model through Cluster Perturbation Theory}},\ }\href
  {https://doi.org/10.1103/PhysRevLett.84.522} {\bibfield  {journal} {\bibinfo
  {journal} {Phys. Rev. Lett.}\ }\textbf {\bibinfo {volume} {84}},\ \bibinfo
  {pages} {522} (\bibinfo {year} {2000})}\BibitemShut {NoStop}%
\bibitem [{\citenamefont {S\'en\'echal}\ \emph {et~al.}(2002)\citenamefont
  {S\'en\'echal}, \citenamefont {Perez},\ and\ \citenamefont
  {Plouffe}}]{PhysRevB.66.075129}%
  \BibitemOpen
  \bibfield  {author} {\bibinfo {author} {\bibfnamefont {D.}~\bibnamefont
  {S\'en\'echal}}, \bibinfo {author} {\bibfnamefont {D.}~\bibnamefont
  {Perez}},\ and\ \bibinfo {author} {\bibfnamefont {D.}~\bibnamefont
  {Plouffe}},\ }\bibfield  {title} {\bibinfo {title} {{Cluster perturbation
  theory for Hubbard models}},\ }\href
  {https://doi.org/10.1103/PhysRevB.66.075129} {\bibfield  {journal} {\bibinfo
  {journal} {Phys. Rev. B}\ }\textbf {\bibinfo {volume} {66}},\ \bibinfo
  {pages} {075129} (\bibinfo {year} {2002})}\BibitemShut {NoStop}%
\bibitem [{\citenamefont {Zhao}\ \emph {et~al.}(2005)\citenamefont {Zhao},
  \citenamefont {Wu},\ and\ \citenamefont {Lin}}]{PhysRevB.71.115201}%
  \BibitemOpen
  \bibfield  {author} {\bibinfo {author} {\bibfnamefont {H.}~\bibnamefont
  {Zhao}}, \bibinfo {author} {\bibfnamefont {C.~Q.}\ \bibnamefont {Wu}},\ and\
  \bibinfo {author} {\bibfnamefont {H.~Q.}\ \bibnamefont {Lin}},\ }\bibfield
  {title} {\bibinfo {title} {{Spectral function of the one-dimensional Holstein
  model at half filling}},\ }\href {https://doi.org/10.1103/PhysRevB.71.115201}
  {\bibfield  {journal} {\bibinfo  {journal} {Phys. Rev. B}\ }\textbf {\bibinfo
  {volume} {71}},\ \bibinfo {pages} {115201} (\bibinfo {year}
  {2005})}\BibitemShut {NoStop}%
\bibitem [{\citenamefont {Ning}\ \emph {et~al.}(2006)\citenamefont {Ning},
  \citenamefont {Zhao}, \citenamefont {Wu},\ and\ \citenamefont
  {Lin}}]{PhysRevLett.96.156402}%
  \BibitemOpen
  \bibfield  {author} {\bibinfo {author} {\bibfnamefont {W.-Q.}\ \bibnamefont
  {Ning}}, \bibinfo {author} {\bibfnamefont {H.}~\bibnamefont {Zhao}}, \bibinfo
  {author} {\bibfnamefont {C.-Q.}\ \bibnamefont {Wu}},\ and\ \bibinfo {author}
  {\bibfnamefont {H.-Q.}\ \bibnamefont {Lin}},\ }\bibfield  {title} {\bibinfo
  {title} {{Phonon Effects on Spin-Charge Separation in One Dimension}},\
  }\href {https://doi.org/10.1103/PhysRevLett.96.156402} {\bibfield  {journal}
  {\bibinfo  {journal} {Phys. Rev. Lett.}\ }\textbf {\bibinfo {volume} {96}},\
  \bibinfo {pages} {156402} (\bibinfo {year} {2006})}\BibitemShut {NoStop}%
\bibitem [{\citenamefont {Yang}\ and\ \citenamefont
  {Feiguin}(2016)}]{PhysRevB.93.081107}%
  \BibitemOpen
  \bibfield  {author} {\bibinfo {author} {\bibfnamefont {C.}~\bibnamefont
  {Yang}}\ and\ \bibinfo {author} {\bibfnamefont {A.~E.}\ \bibnamefont
  {Feiguin}},\ }\bibfield  {title} {\bibinfo {title} {{Spectral function of the
  two-dimensional Hubbard model: A density matrix renormalization group plus
  cluster perturbation theory study}},\ }\href
  {https://doi.org/10.1103/PhysRevB.93.081107} {\bibfield  {journal} {\bibinfo
  {journal} {Phys. Rev. B}\ }\textbf {\bibinfo {volume} {93}},\ \bibinfo
  {pages} {081107} (\bibinfo {year} {2016})}\BibitemShut {NoStop}%
\bibitem [{\citenamefont {Kohno}(2015)}]{PhysRevB.92.085128}%
  \BibitemOpen
  \bibfield  {author} {\bibinfo {author} {\bibfnamefont {M.}~\bibnamefont
  {Kohno}},\ }\bibfield  {title} {\bibinfo {title} {{Spectral properties near
  the Mott transition in the two-dimensional $t\text{\ensuremath{-}}J$
  model}},\ }\href {https://doi.org/10.1103/PhysRevB.92.085128} {\bibfield
  {journal} {\bibinfo  {journal} {Phys. Rev. B}\ }\textbf {\bibinfo {volume}
  {92}},\ \bibinfo {pages} {085128} (\bibinfo {year} {2015})}\BibitemShut
  {NoStop}%
\bibitem [{\citenamefont {Paeckel}\ \emph {et~al.}(2023)\citenamefont
  {Paeckel}, \citenamefont {K{\"o}hler}, \citenamefont {Manmana},\ and\
  \citenamefont {Lenz}}]{paeckel2023matrix}%
  \BibitemOpen
  \bibfield  {author} {\bibinfo {author} {\bibfnamefont {S.}~\bibnamefont
  {Paeckel}}, \bibinfo {author} {\bibfnamefont {T.}~\bibnamefont {K{\"o}hler}},
  \bibinfo {author} {\bibfnamefont {S.~R.}\ \bibnamefont {Manmana}},\ and\
  \bibinfo {author} {\bibfnamefont {B.}~\bibnamefont {Lenz}},\ }\bibfield
  {title} {\bibinfo {title} {{Matrix-product-state-based band-Lanczos solver
  for quantum cluster approaches}},\ }\href
  {https://doi.org/10.48550/arXiv.2310.10799} {\bibfield  {journal} {\bibinfo
  {journal} {arXiv:2310.10799}\ } (\bibinfo {year} {2023})}\BibitemShut
  {NoStop}%
\bibitem [{\citenamefont {Rosenberg}\ \emph {et~al.}(2022)\citenamefont
  {Rosenberg}, \citenamefont {S\'en\'echal}, \citenamefont {Tremblay},\ and\
  \citenamefont {Charlebois}}]{PhysRevB.106.245132}%
  \BibitemOpen
  \bibfield  {author} {\bibinfo {author} {\bibfnamefont {P.}~\bibnamefont
  {Rosenberg}}, \bibinfo {author} {\bibfnamefont {D.}~\bibnamefont
  {S\'en\'echal}}, \bibinfo {author} {\bibfnamefont {A.-M.~S.}\ \bibnamefont
  {Tremblay}},\ and\ \bibinfo {author} {\bibfnamefont {M.}~\bibnamefont
  {Charlebois}},\ }\bibfield  {title} {\bibinfo {title} {{Fermi arcs from
  dynamical variational Monte Carlo}},\ }\href
  {https://doi.org/10.1103/PhysRevB.106.245132} {\bibfield  {journal} {\bibinfo
   {journal} {Phys. Rev. B}\ }\textbf {\bibinfo {volume} {106}},\ \bibinfo
  {pages} {245132} (\bibinfo {year} {2022})}\BibitemShut {NoStop}%
\bibitem [{\citenamefont {Huang}\ \emph {et~al.}(2022)\citenamefont {Huang},
  \citenamefont {Ding}, \citenamefont {Liu},\ and\ \citenamefont
  {Wang}}]{PhysRevResearch.4.L042015}%
  \BibitemOpen
  \bibfield  {author} {\bibinfo {author} {\bibfnamefont {E.~W.}\ \bibnamefont
  {Huang}}, \bibinfo {author} {\bibfnamefont {S.}~\bibnamefont {Ding}},
  \bibinfo {author} {\bibfnamefont {J.}~\bibnamefont {Liu}},\ and\ \bibinfo
  {author} {\bibfnamefont {Y.}~\bibnamefont {Wang}},\ }\bibfield  {title}
  {\bibinfo {title} {{Determinantal quantum Monte Carlo solver for cluster
  perturbation theory}},\ }\href
  {https://doi.org/10.1103/PhysRevResearch.4.L042015} {\bibfield  {journal}
  {\bibinfo  {journal} {Phys. Rev. Res.}\ }\textbf {\bibinfo {volume} {4}},\
  \bibinfo {pages} {L042015} (\bibinfo {year} {2022})}\BibitemShut {NoStop}%
\bibitem [{\citenamefont {Dargel}\ \emph {et~al.}(2012)\citenamefont {Dargel},
  \citenamefont {W\"ollert}, \citenamefont {Honecker}, \citenamefont
  {McCulloch}, \citenamefont {Schollw\"ock},\ and\ \citenamefont
  {Pruschke}}]{PhysRevB.85.205119}%
  \BibitemOpen
  \bibfield  {author} {\bibinfo {author} {\bibfnamefont {P.~E.}\ \bibnamefont
  {Dargel}}, \bibinfo {author} {\bibfnamefont {A.}~\bibnamefont {W\"ollert}},
  \bibinfo {author} {\bibfnamefont {A.}~\bibnamefont {Honecker}}, \bibinfo
  {author} {\bibfnamefont {I.~P.}\ \bibnamefont {McCulloch}}, \bibinfo {author}
  {\bibfnamefont {U.}~\bibnamefont {Schollw\"ock}},\ and\ \bibinfo {author}
  {\bibfnamefont {T.}~\bibnamefont {Pruschke}},\ }\bibfield  {title} {\bibinfo
  {title} {Lanczos algorithm with matrix product states for dynamical
  correlation functions},\ }\href {https://doi.org/10.1103/PhysRevB.85.205119}
  {\bibfield  {journal} {\bibinfo  {journal} {Phys. Rev. B}\ }\textbf {\bibinfo
  {volume} {85}},\ \bibinfo {pages} {205119} (\bibinfo {year}
  {2012})}\BibitemShut {NoStop}%
\bibitem [{\citenamefont {Baker}\ \emph {et~al.}(2021)\citenamefont {Baker},
  \citenamefont {Foley},\ and\ \citenamefont
  {S{\'e}n{\'e}chal}}]{baker2021direct}%
  \BibitemOpen
  \bibfield  {author} {\bibinfo {author} {\bibfnamefont {T.~E.}\ \bibnamefont
  {Baker}}, \bibinfo {author} {\bibfnamefont {A.}~\bibnamefont {Foley}},\ and\
  \bibinfo {author} {\bibfnamefont {D.}~\bibnamefont {S{\'e}n{\'e}chal}},\
  }\bibfield  {title} {\bibinfo {title} {{Direct solution of multiple
  excitations in a matrix product state with block Lanczos}},\ }\href
  {https://doi.org/10.48550/arXiv.2109.08181} {\bibfield  {journal} {\bibinfo
  {journal} {arXiv:2109.08181}\ } (\bibinfo {year} {2021})}\BibitemShut
  {NoStop}%
\bibitem [{\citenamefont {Johnston}\ \emph {et~al.}(2013)\citenamefont
  {Johnston}, \citenamefont {Nowadnick}, \citenamefont {Kung}, \citenamefont
  {Moritz}, \citenamefont {Scalettar},\ and\ \citenamefont
  {Devereaux}}]{PhysRevB.87.235133}%
  \BibitemOpen
  \bibfield  {author} {\bibinfo {author} {\bibfnamefont {S.}~\bibnamefont
  {Johnston}}, \bibinfo {author} {\bibfnamefont {E.~A.}\ \bibnamefont
  {Nowadnick}}, \bibinfo {author} {\bibfnamefont {Y.~F.}\ \bibnamefont {Kung}},
  \bibinfo {author} {\bibfnamefont {B.}~\bibnamefont {Moritz}}, \bibinfo
  {author} {\bibfnamefont {R.~T.}\ \bibnamefont {Scalettar}},\ and\ \bibinfo
  {author} {\bibfnamefont {T.~P.}\ \bibnamefont {Devereaux}},\ }\bibfield
  {title} {\bibinfo {title} {Determinant quantum monte carlo study of the
  two-dimensional single-band hubbard-holstein model},\ }\href
  {https://doi.org/10.1103/PhysRevB.87.235133} {\bibfield  {journal} {\bibinfo
  {journal} {Phys. Rev. B}\ }\textbf {\bibinfo {volume} {87}},\ \bibinfo
  {pages} {235133} (\bibinfo {year} {2013})}\BibitemShut {NoStop}%
\bibitem [{\citenamefont {Mendl}\ \emph {et~al.}(2017)\citenamefont {Mendl},
  \citenamefont {Nowadnick}, \citenamefont {Huang}, \citenamefont {Johnston},
  \citenamefont {Moritz},\ and\ \citenamefont
  {Devereaux}}]{PhysRevB.96.205141}%
  \BibitemOpen
  \bibfield  {author} {\bibinfo {author} {\bibfnamefont {C.~B.}\ \bibnamefont
  {Mendl}}, \bibinfo {author} {\bibfnamefont {E.~A.}\ \bibnamefont
  {Nowadnick}}, \bibinfo {author} {\bibfnamefont {E.~W.}\ \bibnamefont
  {Huang}}, \bibinfo {author} {\bibfnamefont {S.}~\bibnamefont {Johnston}},
  \bibinfo {author} {\bibfnamefont {B.}~\bibnamefont {Moritz}},\ and\ \bibinfo
  {author} {\bibfnamefont {T.~P.}\ \bibnamefont {Devereaux}},\ }\bibfield
  {title} {\bibinfo {title} {{Doping dependence of ordered phases and emergent
  quasiparticles in the doped Hubbard-Holstein model}},\ }\href
  {https://doi.org/10.1103/PhysRevB.96.205141} {\bibfield  {journal} {\bibinfo
  {journal} {Phys. Rev. B}\ }\textbf {\bibinfo {volume} {96}},\ \bibinfo
  {pages} {205141} (\bibinfo {year} {2017})}\BibitemShut {NoStop}%
\bibitem [{\citenamefont {Zhao}\ \emph {et~al.}(2023)\citenamefont {Zhao},
  \citenamefont {Ding},\ and\ \citenamefont {Yang}}]{PhysRevResearch.5.023026}%
  \BibitemOpen
  \bibfield  {author} {\bibinfo {author} {\bibfnamefont {P.-Y.}\ \bibnamefont
  {Zhao}}, \bibinfo {author} {\bibfnamefont {K.}~\bibnamefont {Ding}},\ and\
  \bibinfo {author} {\bibfnamefont {S.}~\bibnamefont {Yang}},\ }\bibfield
  {title} {\bibinfo {title} {{Chebyshev pseudosite matrix product state
  approach for the spectral functions of electron-phonon coupling systems}},\
  }\href {https://doi.org/10.1103/PhysRevResearch.5.023026} {\bibfield
  {journal} {\bibinfo  {journal} {Phys. Rev. Res.}\ }\textbf {\bibinfo {volume}
  {5}},\ \bibinfo {pages} {023026} (\bibinfo {year} {2023})}\BibitemShut
  {NoStop}%
\bibitem [{\citenamefont {Holzner}\ \emph {et~al.}(2011)\citenamefont
  {Holzner}, \citenamefont {Weichselbaum}, \citenamefont {McCulloch},
  \citenamefont {Schollw\"ock},\ and\ \citenamefont {von
  Delft}}]{PhysRevB.83.195115}%
  \BibitemOpen
  \bibfield  {author} {\bibinfo {author} {\bibfnamefont {A.}~\bibnamefont
  {Holzner}}, \bibinfo {author} {\bibfnamefont {A.}~\bibnamefont
  {Weichselbaum}}, \bibinfo {author} {\bibfnamefont {I.~P.}\ \bibnamefont
  {McCulloch}}, \bibinfo {author} {\bibfnamefont {U.}~\bibnamefont
  {Schollw\"ock}},\ and\ \bibinfo {author} {\bibfnamefont {J.}~\bibnamefont
  {von Delft}},\ }\bibfield  {title} {\bibinfo {title} {{Chebyshev matrix
  product state approach for spectral functions}},\ }\href
  {https://doi.org/10.1103/PhysRevB.83.195115} {\bibfield  {journal} {\bibinfo
  {journal} {Phys. Rev. B}\ }\textbf {\bibinfo {volume} {83}},\ \bibinfo
  {pages} {195115} (\bibinfo {year} {2011})}\BibitemShut {NoStop}%
\bibitem [{\citenamefont {Wolf}\ \emph {et~al.}(2015)\citenamefont {Wolf},
  \citenamefont {Justiniano}, \citenamefont {McCulloch},\ and\ \citenamefont
  {Schollw\"ock}}]{PhysRevB.91.115144}%
  \BibitemOpen
  \bibfield  {author} {\bibinfo {author} {\bibfnamefont {F.~A.}\ \bibnamefont
  {Wolf}}, \bibinfo {author} {\bibfnamefont {J.~A.}\ \bibnamefont
  {Justiniano}}, \bibinfo {author} {\bibfnamefont {I.~P.}\ \bibnamefont
  {McCulloch}},\ and\ \bibinfo {author} {\bibfnamefont {U.}~\bibnamefont
  {Schollw\"ock}},\ }\bibfield  {title} {\bibinfo {title} {Spectral functions
  and time evolution from the chebyshev recursion},\ }\href
  {https://doi.org/10.1103/PhysRevB.91.115144} {\bibfield  {journal} {\bibinfo
  {journal} {Phys. Rev. B}\ }\textbf {\bibinfo {volume} {91}},\ \bibinfo
  {pages} {115144} (\bibinfo {year} {2015})}\BibitemShut {NoStop}%
\bibitem [{\citenamefont {Halimeh}\ \emph {et~al.}(2015)\citenamefont
  {Halimeh}, \citenamefont {Kolley},\ and\ \citenamefont
  {McCulloch}}]{PhysRevB.92.115130}%
  \BibitemOpen
  \bibfield  {author} {\bibinfo {author} {\bibfnamefont {J.~C.}\ \bibnamefont
  {Halimeh}}, \bibinfo {author} {\bibfnamefont {F.}~\bibnamefont {Kolley}},\
  and\ \bibinfo {author} {\bibfnamefont {I.~P.}\ \bibnamefont {McCulloch}},\
  }\bibfield  {title} {\bibinfo {title} {Chebyshev matrix product state
  approach for time evolution},\ }\href
  {https://doi.org/10.1103/PhysRevB.92.115130} {\bibfield  {journal} {\bibinfo
  {journal} {Phys. Rev. B}\ }\textbf {\bibinfo {volume} {92}},\ \bibinfo
  {pages} {115130} (\bibinfo {year} {2015})}\BibitemShut {NoStop}%
\bibitem [{\citenamefont {Xie}\ \emph {et~al.}(2018)\citenamefont {Xie},
  \citenamefont {Huang}, \citenamefont {Han}, \citenamefont {Yan},
  \citenamefont {Zhao}, \citenamefont {Xie}, \citenamefont {Liao},\ and\
  \citenamefont {Xiang}}]{PhysRevB.97.075111}%
  \BibitemOpen
  \bibfield  {author} {\bibinfo {author} {\bibfnamefont {H.~D.}\ \bibnamefont
  {Xie}}, \bibinfo {author} {\bibfnamefont {R.~Z.}\ \bibnamefont {Huang}},
  \bibinfo {author} {\bibfnamefont {X.~J.}\ \bibnamefont {Han}}, \bibinfo
  {author} {\bibfnamefont {X.}~\bibnamefont {Yan}}, \bibinfo {author}
  {\bibfnamefont {H.~H.}\ \bibnamefont {Zhao}}, \bibinfo {author}
  {\bibfnamefont {Z.~Y.}\ \bibnamefont {Xie}}, \bibinfo {author} {\bibfnamefont
  {H.~J.}\ \bibnamefont {Liao}},\ and\ \bibinfo {author} {\bibfnamefont
  {T.}~\bibnamefont {Xiang}},\ }\bibfield  {title} {\bibinfo {title}
  {Reorthonormalization of chebyshev matrix product states for dynamical
  correlation functions},\ }\href {https://doi.org/10.1103/PhysRevB.97.075111}
  {\bibfield  {journal} {\bibinfo  {journal} {Phys. Rev. B}\ }\textbf {\bibinfo
  {volume} {97}},\ \bibinfo {pages} {075111} (\bibinfo {year}
  {2018})}\BibitemShut {NoStop}%
\bibitem [{\citenamefont {Jeckelmann}\ and\ \citenamefont
  {White}(1998)}]{PhysRevB.57.6376}%
  \BibitemOpen
  \bibfield  {author} {\bibinfo {author} {\bibfnamefont {E.}~\bibnamefont
  {Jeckelmann}}\ and\ \bibinfo {author} {\bibfnamefont {S.~R.}\ \bibnamefont
  {White}},\ }\bibfield  {title} {\bibinfo {title} {{Density-matrix
  renormalization-group study of the polaron problem in the Holstein model}},\
  }\href {https://doi.org/10.1103/PhysRevB.57.6376} {\bibfield  {journal}
  {\bibinfo  {journal} {Phys. Rev. B}\ }\textbf {\bibinfo {volume} {57}},\
  \bibinfo {pages} {6376} (\bibinfo {year} {1998})}\BibitemShut {NoStop}%
\bibitem [{\citenamefont {Schollw{\"o}ck}(2011)}]{schollwock2011density}%
  \BibitemOpen
  \bibfield  {author} {\bibinfo {author} {\bibfnamefont {U.}~\bibnamefont
  {Schollw{\"o}ck}},\ }\bibfield  {title} {\bibinfo {title} {{The
  density-matrix renormalization group in the age of matrix product states}},\
  }\href {https://doi.org/https://doi.org/10.1016/j.aop.2010.09.012} {\bibfield
   {journal} {\bibinfo  {journal} {Annals of physics}\ }\textbf {\bibinfo
  {volume} {326}},\ \bibinfo {pages} {96} (\bibinfo {year} {2011})}\BibitemShut
  {NoStop}%
\bibitem [{\citenamefont {White}(1992)}]{PhysRevLett.69.2863}%
  \BibitemOpen
  \bibfield  {author} {\bibinfo {author} {\bibfnamefont {S.~R.}\ \bibnamefont
  {White}},\ }\bibfield  {title} {\bibinfo {title} {{Density matrix formulation
  for quantum renormalization groups}},\ }\href
  {https://doi.org/10.1103/PhysRevLett.69.2863} {\bibfield  {journal} {\bibinfo
   {journal} {Phys. Rev. Lett.}\ }\textbf {\bibinfo {volume} {69}},\ \bibinfo
  {pages} {2863} (\bibinfo {year} {1992})}\BibitemShut {NoStop}%
\bibitem [{\citenamefont {White}(1993)}]{PhysRevB.48.10345}%
  \BibitemOpen
  \bibfield  {author} {\bibinfo {author} {\bibfnamefont {S.~R.}\ \bibnamefont
  {White}},\ }\bibfield  {title} {\bibinfo {title} {Density-matrix algorithms
  for quantum renormalization groups},\ }\href
  {https://doi.org/10.1103/PhysRevB.48.10345} {\bibfield  {journal} {\bibinfo
  {journal} {Phys. Rev. B}\ }\textbf {\bibinfo {volume} {48}},\ \bibinfo
  {pages} {10345} (\bibinfo {year} {1993})}\BibitemShut {NoStop}%
\bibitem [{\citenamefont {White}(2005)}]{PhysRevB.72.180403}%
  \BibitemOpen
  \bibfield  {author} {\bibinfo {author} {\bibfnamefont {S.~R.}\ \bibnamefont
  {White}},\ }\bibfield  {title} {\bibinfo {title} {{Density matrix
  renormalization group algorithms with a single center site}},\ }\href
  {https://doi.org/10.1103/PhysRevB.72.180403} {\bibfield  {journal} {\bibinfo
  {journal} {Phys. Rev. B}\ }\textbf {\bibinfo {volume} {72}},\ \bibinfo
  {pages} {180403} (\bibinfo {year} {2005})}\BibitemShut {NoStop}%
\bibitem [{\citenamefont {Wei\ss{}e}\ \emph {et~al.}(2006)\citenamefont
  {Wei\ss{}e}, \citenamefont {Wellein}, \citenamefont {Alvermann},\ and\
  \citenamefont {Fehske}}]{RevModPhys.78.275}%
  \BibitemOpen
  \bibfield  {author} {\bibinfo {author} {\bibfnamefont {A.}~\bibnamefont
  {Wei\ss{}e}}, \bibinfo {author} {\bibfnamefont {G.}~\bibnamefont {Wellein}},
  \bibinfo {author} {\bibfnamefont {A.}~\bibnamefont {Alvermann}},\ and\
  \bibinfo {author} {\bibfnamefont {H.}~\bibnamefont {Fehske}},\ }\bibfield
  {title} {\bibinfo {title} {{The kernel polynomial method}},\ }\href
  {https://doi.org/10.1103/RevModPhys.78.275} {\bibfield  {journal} {\bibinfo
  {journal} {Rev. Mod. Phys.}\ }\textbf {\bibinfo {volume} {78}},\ \bibinfo
  {pages} {275} (\bibinfo {year} {2006})}\BibitemShut {NoStop}%
\bibitem [{\citenamefont {Goodvin}\ \emph {et~al.}(2006)\citenamefont
  {Goodvin}, \citenamefont {Berciu},\ and\ \citenamefont
  {Sawatzky}}]{PhysRevB.74.245104}%
  \BibitemOpen
  \bibfield  {author} {\bibinfo {author} {\bibfnamefont {G.~L.}\ \bibnamefont
  {Goodvin}}, \bibinfo {author} {\bibfnamefont {M.}~\bibnamefont {Berciu}},\
  and\ \bibinfo {author} {\bibfnamefont {G.~A.}\ \bibnamefont {Sawatzky}},\
  }\bibfield  {title} {\bibinfo {title} {{Green's function of the Holstein
  polaron}},\ }\href {https://doi.org/10.1103/PhysRevB.74.245104} {\bibfield
  {journal} {\bibinfo  {journal} {Phys. Rev. B}\ }\textbf {\bibinfo {volume}
  {74}},\ \bibinfo {pages} {245104} (\bibinfo {year} {2006})}\BibitemShut
  {NoStop}%
\bibitem [{\citenamefont {Hohenadler}\ \emph {et~al.}(2003)\citenamefont
  {Hohenadler}, \citenamefont {Aichhorn},\ and\ \citenamefont {von~der
  Linden}}]{PhysRevB.68.184304}%
  \BibitemOpen
  \bibfield  {author} {\bibinfo {author} {\bibfnamefont {M.}~\bibnamefont
  {Hohenadler}}, \bibinfo {author} {\bibfnamefont {M.}~\bibnamefont
  {Aichhorn}},\ and\ \bibinfo {author} {\bibfnamefont {W.}~\bibnamefont
  {von~der Linden}},\ }\bibfield  {title} {\bibinfo {title} {{Spectral function
  of electron-phonon models by cluster perturbation theory}},\ }\href
  {https://doi.org/10.1103/PhysRevB.68.184304} {\bibfield  {journal} {\bibinfo
  {journal} {Phys. Rev. B}\ }\textbf {\bibinfo {volume} {68}},\ \bibinfo
  {pages} {184304} (\bibinfo {year} {2003})}\BibitemShut {NoStop}%
\bibitem [{\citenamefont {Kohno}(2018)}]{kohno2018characteristics}%
  \BibitemOpen
  \bibfield  {author} {\bibinfo {author} {\bibfnamefont {M.}~\bibnamefont
  {Kohno}},\ }\bibfield  {title} {\bibinfo {title} {Characteristics of the mott
  transition and electronic states of high-temperature cuprate superconductors
  from the perspective of the hubbard model},\ }\href
  {https://doi.org/10.1088/1361-6633/aaa53d} {\bibfield  {journal} {\bibinfo
  {journal} {Rep. Prog. Phys.}\ }\textbf {\bibinfo {volume} {81}},\ \bibinfo
  {pages} {042501} (\bibinfo {year} {2018})}\BibitemShut {NoStop}%
\bibitem [{\citenamefont {Wang}\ \emph {et~al.}(2020)\citenamefont {Wang},
  \citenamefont {He}, \citenamefont {Wohlfeld}, \citenamefont {Hashimoto},
  \citenamefont {Huang}, \citenamefont {Lu}, \citenamefont {Mo}, \citenamefont
  {Komiya}, \citenamefont {Jia}, \citenamefont {Moritz} \emph
  {et~al.}}]{wang2020emergence}%
  \BibitemOpen
  \bibfield  {author} {\bibinfo {author} {\bibfnamefont {Y.}~\bibnamefont
  {Wang}}, \bibinfo {author} {\bibfnamefont {Y.}~\bibnamefont {He}}, \bibinfo
  {author} {\bibfnamefont {K.}~\bibnamefont {Wohlfeld}}, \bibinfo {author}
  {\bibfnamefont {M.}~\bibnamefont {Hashimoto}}, \bibinfo {author}
  {\bibfnamefont {E.~W.}\ \bibnamefont {Huang}}, \bibinfo {author}
  {\bibfnamefont {D.}~\bibnamefont {Lu}}, \bibinfo {author} {\bibfnamefont
  {S.-K.}\ \bibnamefont {Mo}}, \bibinfo {author} {\bibfnamefont
  {S.}~\bibnamefont {Komiya}}, \bibinfo {author} {\bibfnamefont
  {C.}~\bibnamefont {Jia}}, \bibinfo {author} {\bibfnamefont {B.}~\bibnamefont
  {Moritz}}, \emph {et~al.},\ }\bibfield  {title} {\bibinfo {title} {Emergence
  of quasiparticles in a doped mott insulator},\ }\href
  {https://doi.org/https://doi.org/10.1038/s42005-020-00480-5} {\bibfield
  {journal} {\bibinfo  {journal} {Commun. Phys.}\ }\textbf {\bibinfo {volume}
  {3}},\ \bibinfo {pages} {210} (\bibinfo {year} {2020})}\BibitemShut {NoStop}%
\bibitem [{\citenamefont {Kohno}(2012)}]{PhysRevLett.108.076401}%
  \BibitemOpen
  \bibfield  {author} {\bibinfo {author} {\bibfnamefont {M.}~\bibnamefont
  {Kohno}},\ }\bibfield  {title} {\bibinfo {title} {{Mott Transition in the
  Two-Dimensional Hubbard Model}},\ }\href
  {https://doi.org/10.1103/PhysRevLett.108.076401} {\bibfield  {journal}
  {\bibinfo  {journal} {Phys. Rev. Lett.}\ }\textbf {\bibinfo {volume} {108}},\
  \bibinfo {pages} {076401} (\bibinfo {year} {2012})}\BibitemShut {NoStop}%
\bibitem [{\citenamefont {Braun}\ and\ \citenamefont
  {Schmitteckert}(2014)}]{PhysRevB.90.165112}%
  \BibitemOpen
  \bibfield  {author} {\bibinfo {author} {\bibfnamefont {A.}~\bibnamefont
  {Braun}}\ and\ \bibinfo {author} {\bibfnamefont {P.}~\bibnamefont
  {Schmitteckert}},\ }\bibfield  {title} {\bibinfo {title} {{Numerical
  evaluation of Green's functions based on the Chebyshev expansion}},\ }\href
  {https://doi.org/10.1103/PhysRevB.90.165112} {\bibfield  {journal} {\bibinfo
  {journal} {Phys. Rev. B}\ }\textbf {\bibinfo {volume} {90}},\ \bibinfo
  {pages} {165112} (\bibinfo {year} {2014})}\BibitemShut {NoStop}%
\end{thebibliography}%
\end{document}